\documentclass[showpacs,preprintnumbers,amsmath,amssymb,10pt]{revtex4}

\usepackage{amsfonts,amssymb,amscd,amsmath}
\usepackage{graphicx}

 \newcommand\beq{\begin{equation}}
 \newcommand\eeq{\end{equation}}
 \newcommand\beqn{\begin{eqnarray}}
 \newcommand\eeqn{\end{eqnarray}}
 \newcommand{\la}{\langle}
 \newcommand{\ra}{\rangle}
 
\def\fm{\,\mbox{fm}}
\def\GeV{\,\mbox{GeV}}

 \def\Pom{{ I\!\!P}}

\def\im{{\rm Im}}

\def\fm{\,\mbox{fm}}
\def\GeV{\,\mbox{GeV}}

\def\Pom{{\bf I\!P}}

\begin{document}

\title{Azimuthal Asymmetry of pions in $pp$ and $pA$ collisions}
\author{B. Z. Kopeliovich}
\author{A. H. Rezaeian}
\author{Ivan Schmidt}
\affiliation{Departamento de F\'\i sica y Centro de Estudios Subat\'omicos, Universidad
T\'ecnica Federico Santa Mar\'\i a, Casilla 110-V, Valpara\'\i so,
Chile}

\begin{abstract}
We investigate the azimuthal asymmetry v2 of produced pions in $pp$
and $pA$ collisions at both RHIC and SPS energies. In our approach, based on the pQCD
parton model and the light-cone QCD-dipole formalism, the azimuthal
asymmetry results from a correlation between the color-dipole
orientation and the impact parameter of the collision. We
introduce the color-dipole orientation within an improved Born
approximation and the saturation model which satisfies
available DIS data, showing that the azimuthal asymmetry of partons
and pions is very sensitive to the choice of the model, and that it  is
reduced in the saturation model. We find that v2 of
quarks and gluons in parton-nucleus collisions have very different
patterns. The azimuthal asymmetry of gluons in gluon-nucleus collisions
can be negative at small transverse momentum, changes the sign and
becomes positive at high transverse momentum. The azimuthal asymmetry of quarks in
quark-nucleus collisions is positive at all values of transverse momentum. We
find that the azimuthal anisotropy v2 of produced pions in both pp
and pA collisions is positive, albeit rather small.
\end{abstract}

\maketitle
\date{\today}

\section{Introduction}

One of the most important discoveries in ultrarelativistic nucleus-nucleus collisions at the Relativistic
Heavy Ion Collider (RHIC) was the observation of a large
elliptic flow \cite{rhic1}, which is as large as the one
predicted by ideal-fluid models (which assume zero viscosity) \cite{rhic2,hadro}.
Microscopically, the elliptic flow results from the interactions
between the produced particles, and therefore carries information about
the dense matter produced at RHIC.

Various theoretical calculations \cite{vrhic} support the notion that
collective flow is perhaps generated early in the nucleus-nucleus
collision, and is present at the partonic level, before partons
coalesce or the hadronic fragmentation stage. On the other hand, it has
been shown that two-body interactions between partons cannot by themselves
generate sufficient flow to explain the observations, unless partonic
cross sections are artificially enhanced by more than an order of
magnitude over perturbative QCD predictions \cite{vrhic2}, (see, however, the recent development on this line in Ref.~\cite{vvv}).  This
indicates that the quark-gluon matter created at RHIC is strongly
interacting, unlike the type of weakly interacting quark-gluon plasma
expected to occur at very high temperatures on the basis of asymptotic
freedom.  Ultrarelativistic heavy ion collisions probe QCD in different regimes at different stages of the collisions.
To be able to understand the underlying  dynamics of the collisions and in particular the observed elliptic flow, one has to learn how to
disentangle between the contributions of different physical subprocesses and their contributions to the observed azimuthal asymmetry.


We have proposed recently a scenario which produces an
azimuthal asymmetry coming from the initial stage of relativistic
nuclear collisions \cite{me1,me2}. In our approach, the azimuthal
asymmetry is related to the sensitivity of parton multiple
interactions to the steep variation of the nuclear density at the edge
of the nuclei via the color dipole orientation. This effect is also
present in elementary $pp$ reactions where the correlation between the
color-dipole orientation and the impact parameter of the collision
leads to an azimuthal asymmetry. We have recently computed the azimuthal
asymmetry of the prompt photons in nuclear collisions coming from this
mechanism \cite{me2}.

In this paper we use this idea to calculate the azimuthal asymmetry of pions produced in
$pp$ and $pA$ collisions. Measuring $v_2$ of
the produced particles in $pp$ and $pA$ collisions is a challenge for
experimentalists, partly due to the difficulties associated with the identification of the reaction
plane. Nevertheless, measurements of the azimuthal correlations from
$pp$, $dAu$ and $AuAu$ collisions at RHIC indicate that the azimuthal
asymmetries in $pp$ and $pA$ collisions can be quite different from those in
$AA$ collisions \cite{rhic-vpp}.

Here we systematically study the azimuthal asymmetry of
quarks and radiated gluons in parton-nucleus collisions, and show how they
contribute to the azimuthal asymmetry of produced
hadrons in $pp$ and $pA$ collisions. The basic tools of our
calculations include the pQCD parton model combined with the light-cone QCD-dipole
formalism. Such a study helps us to understand how the azimuthal
asymmetry of the produced particles evolves from elementary $pp$
collisions to cold nuclear matter in $pA$ collisions and then to
a hot quark-gluon plasma in $AA$ collisions.  The azimuthal asymmetry
of the produced particles in peripheral $AA$ collisions or at high
$p_T$ is expected to be similar to those in $pA$ and $pp$
reactions. Therefore, this study may be also relevant for the physics of
$AA$ reactions and can be used as a baseline for jet-quenching
models, since in $pp$ and $pA$ collisions no hot and dense medium is
created.

This paper is organized as follows: in Sec. I and II we introduce the
color-dipole orientation in an improved Born approximation \cite{zkl} and
within the saturation model of Golec-Biernat and W\"usthoff
\cite{gbw}. In Sec. IV, we calculate the azimuthal asymmetry of quarks
in quark-nucleus collisions. In this section, the broadening of a
projectile parton in $pA$ collisions will be introduced. This is the
key ingredient of the Cronin effect and hadron production in $pA$
collisions. In Sec. V, we study the azimuthal asymmetry of the
radiated gluons in gluon-nucleon and gluon-nucleus collisions. In
Sec. VI, we discuss hadron production in the short- and
long-coherence regimes. We introduce two very different schemes for
hadron production in $pp$ and $pA$ collisions: (1) pQCD-improved
factorization and (2) the light-cone QCD-dipole factorization scheme. The
numerical results and discussions are given in Sec. VII. Some
concluding remarks are given in Sec. VIII. In an Appendix, we give some
details of the calculations carried out in Sec V.

\section{Color dipole orientation in Born approximation}
A colorless $\bar qq$ dipole is able to interact only due to the
difference between the impact parameters of the $q$ and $\bar q$ relative
to the scattering center. Therefore, a $\bar qq$ fluctuation cannot be produced
if both $q$ and $\bar q$ have the same impact
parameter $\vec b$ from the target, even if their transverse separation $r$
is not zero. In terms of the partial elastic amplitude
$f_{q\bar{q}}(\vec{b},\vec{r})$, it means that the vectors
$\vec{r}$ and $\vec b$ are correlated.  This can be seen in a simple example of a dipole interacting with a quark in Born approximation. The partial elastic amplitude up to a factor $\mathcal{N}$ reads,
\beqn
\im f_{\bar qq}(\vec b ,\vec r) &=& \frac{\mathcal{N}}{(2\pi)^2} \int
\frac{d^2q\,d^2q'}{(q^2+m_g^2)(q'^{2}+m_g^2)}\,
\left[e^{i\vec q\cdot(\vec b +\vec r/2)}- e^{i\vec q\cdot(\vec b -\vec
r/2)}\right]
\left[e^{i\vec q^{\,\prime}\cdot(\vec b +\vec r/2)}- e^{i\vec
q^{\,\prime}\cdot(\vec b -\vec r/2)}\right],\nonumber\\
&=& \mathcal{N}
\left[K_0\left(m_g\left|\vec b +\frac{\vec r}{2}\right|\right) -
K_0\left(m_g\left|\vec b -\frac{\vec r}{2}\right|\right)\right]^2,
\hspace{1cm} \text{(2g Model)},\
\label{e1}
 \eeqn
 where $K_0(x)$ is the modified Bessel function and
we introduced an effective gluon mass $m_g$ to take
 into account some nonperturbative effects. It is obvious from the
above expression that the partial
 elastic dipole amplitude exposes a correlation between $\vec r$ and
 $\vec b $, and the amplitude vanishes when $\vec b \cdot\vec r=0$.


The Born amplitude is unrealistic, since it leads to an energy
independent dipole cross section $\sigma_{\bar qq}(r,x)$ where $x$ corresponds to
the Bjorken variable in DIS. The energy
dependence of the dipole cross section is generated by additional
radiation of gluons, which can be resummed in the leading $ln(1/x)$
approximation. As a first step to improve the above approximation, we first obtain
the dipole cross section from Eq.~(\ref{e1}) at small $r$ \cite{zkl},
\beq
\lim_{r\to 0}\sigma_{q\bar{q}}(r,x)=2\int d^{2}\vec b~\text{Im}f_{q\bar{q}}(\vec b,\vec{r})
\approx \mathcal{N} 2\pi \left(0.62-\ln(m_g r)\right) r^2,
\eeq
the unknown coefficient $\mathcal{N}$ can be then obtained by comparing
the above expression with the small $r$ expansion of the Golec-Biernat
and W\"usthoff (GBW) dipole cross section \cite{gbw}. Notice that at small $r$
there is no logarithmic term in the GBW dipole cross section, therefore we fix the mean value of $\bar r$,  which depends on the process under consideration.
Thus, we obtain
\beq
\mathcal{N}= \frac{\sigma_{0}}{2\pi\left(0.62-\ln(m_g \bar r)\right) R_{0}^2(x)},
\eeq
where $\sigma_{0}=23.03$ mb, $R_{0}(x)=0.4 \fm \times
(x/x_{0})^{0.144}$ with $x_{0}=3.04\times 10^{-4}$. Notice that the
amplitude given in Eq.~(\ref{e1}) is for the $\bar qq$ dipole colliding with a quark target, not with a
nucleon. Although by fixing the coefficient $\mathcal{N}$ with the GBW
cross section some of the missing effects are incorporated,  still the
above approximation is rather crude. In the next section, we introduce the dipole orientation without using any approximation.

\section{Color dipole orientation in the saturation model}
Here, we introduce the color dipole orientation within
the phenomenological saturation model of GBW, which includes contributions from higher order perturbative
corrections as well as non-perturbative effects contained in DIS data.  The dipole elastic amplitude $f^{N}_{q\bar{q}}$ of a $\bar{q}q$ dipole
colliding with a proton at impact parameter $\vec b$ is given by \cite{me1}
 \beqn
&&\im f^N_{\bar qq}(\vec b ,\vec r,x,\beta)=\frac{1}{12\pi}
\int\frac{d^2q\,d^2q'}{q^2\,q'^2}\,\alpha_s\,
{\cal F}(x,\vec q,\vec q^{\,\prime})
e^{i\vec b \cdot(\vec q-\vec q^{\,\prime})}
\left(e^{-i\vec q\cdot\vec r\beta}-
e^{i\vec q\cdot\vec r(1-\beta)}\right)\,
\left(e^{i\vec q'\cdot\vec r\beta}-
e^{-i\vec q'\cdot\vec r(1-\beta)}\right)\,
\,,
\label{300}
 \eeqn
 where we defined
 $\alpha_{s}=\sqrt{\alpha_{s}(q^{2})\alpha_{s}(q^{\prime 2})}$, and
 $\mathcal{F}(x,\vec{q},\vec{q}^{\,\prime})$ is the generalized
 unintegrated gluon density. The fractional light-cone momenta of the
 quark and antiquark are denoted by $\beta$ and $1-\beta$,
 respectively.  In the 2g model, Eq.~(\ref{e1}),  we assumed that
that $q$ and $\bar{q}$  have equal longitudinal momenta, i.e. they
are equally distant from the dipole center of gravity, which
corresponds to the case with a parameter $\beta=1/2$. The
generalized unintegrated gluon density was proposed in Ref.
\cite{me1} assuming that the transverse momentum distributions of
the two gluons do not correlate, except for the
 Pomeron-proton vertex, which is function of the total momentum transfer, and it is given by
\beq {\cal
F}(x,\vec q,\vec q^{\,\prime})=
\frac{3\,\sigma_0}{16\,\pi^2\,\alpha_s}\ q^2\,q'^2\,R_0^2(x)
{\rm exp}\Bigl[-{1\over8}\,R_0^2(x)\,(q^2+q'^2)\Bigr]
{\rm exp}\bigl[-R^2_N(x)(\vec q-\vec q^{\,\prime})^2/2\bigr]
\,,
 \label{320} \eeq
 This generalized unintegrated gluon density is related to the diagonal one by
 $\mathcal{F}(x,\vec{q},\vec{q}^{\,\prime}=\vec{q})=\mathcal{F}(x,q)$. We
assume here that the transverse momentum dependence of the
dipole-proton elastic amplitude has a Gaussian form. Comparison with
the saturated form \cite{gbw} of the dipole-proton cross section
$\sigma_{q\bar{q}}(r,x)$, \beqn
\sigma_{q\bar{q}}(r,x)&=&2\int d^{2}\vec b~\text{Im}f^{N}_{q\bar{q}}(\vec b,\vec{r},x,\beta),\label{di-app0}  \\
&=&\frac{4\pi}{3}\int\frac{d^{2}q}{q^{4}}(1-e^{-i\vec{q}.\vec{r}})\alpha_{s}(q^{2})\mathcal{F}(x,q),
\label{di-app}\
\eeqn
fixes the form of ${\cal F}(x,\vec q,\vec q^{\,\prime})$ and the values of the parameters, except for $R_N^2(x)$.

Knowing the generalized unintegrated gluon density, we can now
perform the integration in Eq.~(\ref{300}) and obtain the partial
elastic dipole-proton amplitude, \beqn \im f^N_{\bar qq}(\vec b,\vec
r,x,\beta) &=& \frac{\sigma_0}{8\pi \mathcal{B}(x)}\,
\Biggl\{\exp\left[-\frac{[\vec b+\vec
r(1-\beta)]^2}{2\mathcal{B}(x)}\right] + \exp\left[-\frac{(\vec
b-\vec r\beta)^2}{2\mathcal{B}(x)}\right]
-2\exp\Biggl[-\frac{r^2}{R_0^2(x)}- \frac{[\vec b+(1/2-\beta)\vec
r]^2}{2\mathcal{B}(x)}\Biggr]
\Biggr\}, \nonumber\\
&& \hspace{10.5cm} \text{(GBW model)}\
\label{340}
 \eeqn
with the notation $\mathcal{B}(x)=R_N^2(x)+R_0^2(x)/8$.

To fix the function $R_N^2(x)$ we use another observable, the $t$-slope $B^{\bar qq}_{el}(x,r)$ of the elastic dipole-proton cross section (at $t=0$) in the limit of vanishingly small dipole, $r\to0$. In this limit,
\beqn
B^{\bar qq}_{el}(x)&=&{1\over2}\la b^2\ra = \frac{1}{\sigma_{\bar qq}(r)}\int d^2b\,b^2\,
\im f^N_{\bar qq}(\vec b,\vec r,x,\beta=1/2)\ =\mathcal{B}(x)+\frac{r^2}{8\left(1-e^{-r^2/R_0^2(x)}\right)},\nonumber\\
B^{\bar qq}_{el}(x,r\to0)&=&\mathcal{B}(x)+\frac{1}{8}R_0^2(x).
\label{345} \eeqn In this expression we fixed $\beta=1/2$ for the
sake of simplicity. Equation (\ref{345}) can be compared with the
slope of the cross section of elastic electroproduction of
$\rho$-mesons measured at HERA at small $x$ and high $Q^2$. It was
observed that at $Q^2\gg1\GeV^2$ the slope saturates at the value
$B_{\gamma^*p\to\rho p}(x,Q^2\gg1\GeV^2)\approx 5\GeV^{-2}$
\cite{zues}, which can be compared with our result Eq.~(\ref{345})
in the limit $r\to 0$, since at high $Q^2$ the effective size of the
dipole is vanishingly small. Therefore, we have
$R_N^2(x)=B_{\gamma^*p\to\rho
p}(x,Q^2\gg1\GeV^2)-\frac{1}{4}R_0^2(x)$.

Notice that the expression Eq.~(\ref{340}) at $r\to0$ exposes the
property of color transparency \cite{zkl}, $f^N_{\bar qq}(\vec
b,\vec r,x,\beta)\propto r^2$. It also goes beyond the usual
assumption that the dipole cross section is independent of the
light-cone momentum sharing $\beta$. Although, the partial amplitude
Eq.~(\ref{340}) does depend on $\beta$, this dependence disappears
after integration over impact parameter $\vec b $ as shown in
Eq.~(\ref{di-app}). From Eq.~(\ref{340}), it is seen that when the
transverse dipole size $r$ and the impact parameter $b$ become
comparable in size then the orientation becomes important. For very
small $r$ or $b$, the dipole orientation is not present. The partial
dipole amplitude behaviour also changes with the parameter $\beta$
\cite{me3}.

One should note that the GBW model is a simple parametrization
which has some restrictions. In particular, the model exhibits no
power-law tails in momentum space in contradiction with QCD.
Besides, it does not match the QCD
evolution (DGLAP) at large values of $Q^2$. 
Therefore, one should be cautious applying this model at very high transverse
momenta accessible at the energies of LHC.

In the above we relied on the saturation GBW model, which depends on
Bjorken $x$. However, for soft reactions the c.m. energy $s$, rather
than Bjorken $x$, is the proper variable. Similar to the GBW model,
the $s$-dependent dipole cross section with a saturated shape fitted
to data on DIS at $Q^2$ not high, and to real photo-absorption and
photoproduction of vector mesons, was introduced in
Ref.~\cite{kst2},

 \beq
\sigma_{\bar qq}(r,s)=\sigma_0(s)\left[1-e^{-r^2/R_0^2(s)}\right]\,,
\label{300-2}
 \eeq
where $R_0(s)=0.88\,\text{fm}\,(s_0/s)^{0.14}$ with
 $s_0=1000~\text{GeV}^2$. The normalization factor $\sigma_0(s)$ is
 fixed by demanding that the pion-proton total cross section be
 reproduced, that is $\int d^2r\,|\Psi_\pi(r)|^2\sigma_{\bar
 qq}(r,s)=\sigma^{\pi p}_{tot}(s)$, where the pion
wave function squared integrated over longitudinal quark momenta has the
form,
 \beq
\left|\Psi_\pi(\vec r)\right|^2 =
\frac{3}{8\pi \la r^2_{ch}\ra_\pi}
\exp\left(-\frac{3r^2}{8\la r^2_{ch}\ra_\pi}\right)\,,
\label{310-2}
 \eeq
 with a mean pion charge
radius squared $\la r^2_{ch}\ra_\pi=0.44\fm^2$ \cite{r-pion}. In this way, the normalization factor $\sigma_0(s)$ is determined,
 \beq
\sigma_0(s)=\sigma^{\pi p}_{tot}(s)\,
\left(1 + \frac{3\,R^2_0(s)}{8\,\la r^2_{ch}\ra_{\pi}}
\right)\,.
\label{320-2}
 \eeq
We employ the parametrization of the fit in Ref.~\cite{pom} for the
Pomeron part of the cross section $\sigma^{\pi
p}_{tot}(s)=23.6(s/s_0)^{0.08} ~\text{mb}$, where $s_0=1000\,
\text{GeV}^2$.

We assume that for soft processes one can switch from $x$- to
$s$-dependence, keeping the same functional form of the dipole amplitude
Eq.~(\ref{340}) but adjusting the parameters $ R_N^2(s)$ and
$\sigma_0(s)$ to observables in soft reactions. The first condition is
that the $s$-dependent dipole partial amplitude reproduces the $s$-dependent
pion-proton cross section. Another
condition is the reproduction of  the slope  at $t=0$, $B_{el}^{\pi p}(s)={1\over2}\la b^2\ra$.
These conditions will be satisfied by the following replacements:
\beqn
\im f^N_{\bar qq}(\vec b,\vec r,x,\beta)&\Rightarrow&\im f^N_{\bar qq}(\vec b,\vec r,s,\beta),\nonumber\\
R_0(x)&\Rightarrow& R_0(s)=0.88\,\text{fm}\,(s_0/s)^{0.14}\nonumber\\
 R_N^2(x) &\Rightarrow&  R_N^2(s)=B^{\pi p}_{el}(s)-\frac{1}{4}R_0^2(s)-\frac{1}{3}\la r_{ch}^2\ra_\pi,\nonumber\\
\sigma_0&\Rightarrow& \sigma_0(s)=
\sigma^{\pi p}_{tot}(s)\,\left(1 + \frac{3\,R^2_0(s)}{8\,\la r^2_{ch}\ra_{\pi}}
\right),
\hspace{1.5cm} \text{(KST model)}\
\label{sat2}
\eeqn
where we use a Regge parametrization
for the elastic slope, $B^{\pi p}_{el}(s)=B_0+2\alpha_\Pom^\prime
\ln(s/\mu^2)$, with $B_0=6~\text{GeV}^{-2}$,
$\alpha_\Pom^\prime=0.25\GeV^{-2}$, and $\mu^2=1\GeV^2$.
In what follows, we call the $s$-dependent dipole amplitude KST model.

\section{$\bf{v_2}$ of quarks }

Multiple interactions of projectile quarks in the target may proceed
coherently or incoherently. In the former case the multiple
interaction amplitude is a convolution of single scattering
amplitudes, and  in the latter case one should convolute
differential cross sections, rather than amplitudes. The condition
of coherence is exactly the same as in classical optics, namely the
maximal longitudinal distance between different scattering centers
should not considerably exceed the so called coherence length, \beq
l_c=\frac{2E_q}{p_T^2}, \label{coh} \eeq where $E_q$ is the quark
energy in the nuclear rest frame and $p_T$ is the total transverse
momentum of the quark accumulated from the multiple rescatterings.
Notice that at mid rapidities we have $E_q=p_T\sqrt{s}/m_N$. Thus,
one can assume a coherent regime of multiple interactions only for
not very large transverse momenta, which is restricted at mid
rapidities by \beq p_T\lesssim \frac{\sqrt{s}}{m_N R_A}.
\label{pt-coh} \eeq At the RHIC energy, $\sqrt{s}=200\GeV$, only
quarks with up to several GeV transverse momentum can be produced
coherently.

The transverse momentum distribution of a parton after propagation
through a nucleus is given by the square of the amplitude of multiple interactions.
The quark impact parameters in these two amplitudes, direct and conjugated, are different.
As a result, one can express the $p_T$-distribution in terms of the eikonalized partial elastic
$q\bar q$ dipole amplitude \cite{boris0},
\beq
\frac{d \sigma^{q}(qA\to qX) }{d^{2}\vec{p}_{T}d^{2}\vec{b}}(b,p_T,x)=\frac{1}{(2\pi)^{2}}
\int d^{2}\vec{r}_{1}d^{2}\vec{r}_{2}e^{i \vec{p}_{T}.(\vec{r}_{1}-\vec{r}_{2})}\Omega^{q}_{in}
(\vec{r}_{1},\vec{r}_{2})
(1-\text{Im}f^{A}_{q\bar{q}}(b,(\vec{r}_1-\vec{r}_2),\beta)),\label{main}
\eeq
where $\Omega^{q}_{in}(\vec{r}_{1},\vec{r}_{2})$ is the density matrix which
describes the impact parameter distribution of the quark in the
incident hadron,
\begin{equation}
\Omega^{q}_{in}(\vec{r}_{1},\vec{r}_{2})=\frac{\langle k_{T}^{2}\rangle}{\pi}
e^{-\frac{1}{2}(r_{1}^{2}+r_{2}^{2})\langle k_{T}^{2}\rangle}, \label{kkt}
\end{equation}
where $\langle k_{T}^{2}\rangle$ denotes the mean value of the
parton primordial transverse momentum squared. The function
$\text{Im}f^{A}_{q\bar{q}}$ in the above denotes the partial
amplitude of a $q\bar{q}$ dipole colliding with a nucleus at impact
parameter $\vec{b}$ and can be written, in the eikonal form, in
terms of the dipole elastic amplitude $f^{N}_{q\bar{q}}$ of a
$\bar{q}q$ dipole colliding with a proton at impact parameter
$\vec{b}$,
\beq
\text{Im}f^{A}_{q\bar{q}}(\vec{b},\vec{r},\beta)=
1-\exp[-\int d^{2}\vec{s}~
\text{Im}f^{N}_{q\bar{q}}(\vec{s},\vec{r},\beta)T_{A}(\vec{b}+\vec{s})]
\label{eik},
\eeq
where $ T_A(\vec{b})=\int dz
\rho_A(\vec{b},z)$ is the nuclear thickness function and
$\rho_A(\vec{b},z)$ denotes the nuclear density at a impact
parameter $\vec{b}$ and a longitudinal coordinate $z$. The
fractional light cone momentum $x$  of
 the target gluons is implicit in the above expression, and the parameter
 $\beta$ in the $q\bar q$ dipole amplitude of the saturation model is
 taken to be $1/2$.

The integral over $\vec{r_{1}}$ in Eq.~(\ref{main}) can be readily done.  In order to
compute the remaining integrals, we choose a coordinate in which the
angle between the impact parameter $\vec{b}$ with $\vec{s}$, $\vec{r}=\vec{r_{1}}-\vec{r_{2}}$
and $\vec{p}_{T}$ are denoted by $\theta$, $\delta$ and $\phi$,
respectively.  Then equation (\ref{main}) takes the following simple form,
\begin{eqnarray}
\frac{d \sigma^{q}(qA\to qX) }{d^{2}\vec{p}_{T}d^{2}\vec{b}}
&=&
\frac{1}{(2\pi)^2}\int d\beta rdr e^{i p_{T} r \cos(\phi-\delta)-\frac{\langle k_{T}^{2}
\rangle}{4}r^{2}-\mathcal{I}(b,r,\delta)},\nonumber\\
\label{mainv}\
\end{eqnarray}
with the notation,
\begin{eqnarray}
\mathcal{I}(b, r,\delta)&=&
\int d^{2}\vec{s}~\text{Im}f^{N}_{q\bar{q}}(\vec{s},\vec{r})T_{A}(\vec{b}+\vec{s}).
\label{I1}\
\end{eqnarray}
The result of the convolution of the nuclear thickness with the
dipole amplitude will explicitly depend on the angle $\delta$ between the impact parameter $\vec b$ and the
dipole transverse vector $\vec r$. The azimuthal asymmetry resulting from a single quark passing
through the nucleus is computed as a second order Fourier coefficient
in a Fourier expansion of the azimuthal dependence of a
single-particle spectra Eq.~(\ref{main}) around the beam direction,
\begin{equation}
v_{2}^{q}(p_{T},b)= \frac{\int_{-\pi}^{\pi} d\phi \cos(2\phi)  \frac{d \sigma^{q}(qA\to qX) }
{d^{2}\vec{p}_{T}d^{2}\vec{b}}}
{\int_{-\pi}^{\pi} d\phi \frac{d \sigma^{q}(qA\to qX) }{d^{2}\vec{p}_{T}d^{2}\vec{b}}  }. \label{v2-1}
\end{equation}
Substituting the expression in Eq.~(\ref{mainv}) into
Eq.~(\ref{v2-1}), the integral over $\phi$ can then be performed
analytically by using the identity given in Eq.~(\ref{id}),  getting
\begin{equation}
v_{2}^q(p_{T},b)=-\frac{\int_{0}^{2\pi} d\delta \int_{0}^{\infty} rdr
J_{2}(p_{T}r)\cos(2\delta)e^{-\frac{\langle k_{T}^{2}\rangle}{4}r^{2}-\mathcal{I}(b,r,\delta)}}
{\int_{0}^{2\pi} d\delta \int_{0}^{\infty} rdr J_{0}(p_{T}r)e^{-\frac{\langle k_{T}^{2}\rangle}{4}r^{2}-
\mathcal{I}(b,r,\delta)}}.\label{sv2}
\end{equation}
where $J_{n}(x)$ denotes the Bessel function. In the above expression
the angle dependence $\phi$ between the impact parameter $\vec{b}$ and
the transverse momentum of the projectile quark $\vec{p}_{T}$ disappeared
and the azimuthal asymmetry is directly related to the dipole orientation
with respect to impact parameter $\vec{b}$ through the angle
$\delta$. If one neglects the angular dependence of dipole cross section, then $v_{2}^{q} $
becomes identically zero regardless of a given nuclear profile and
of dipole amplitude parametrization.
\section{$\bf{v_2}$ of gluons}

One could calculate the cross section of high-$p_T$ gluon production
in the same way as was done for quarks. Namely gluons can also
experience multiple coherent interactions provided that the final
$p_T$ is restricted by the condition Eq.~(\ref{pt-coh}). One can
neglect interaction with the spectator partons, like we did it for
quarks in the previous section, if the final $p_T$ of the parton is
much larger than its primordial transverse momentum. Otherwise,
interaction with spectators is important since color screening is at
work. Such an approximation is valid for valence quarks, which have a
primordial momentum of the order of $\Lambda_{QCD}$, starting from
$p_T$ of several hundred MeV.

Nevertheless, the dynamics of gluon radiation is more involved.
There are many experimental evidences \cite{kst2,spot} for a much
higher primordial momentum of gluons compared to quarks.  Therefore,
one can neglect spectators only at $p_T$ of several GeV. In order to
be able to work at smaller $p_T$ one should include interaction with
spectators, i.e. instead of "elastic" gluon scattering, $GN\to GX$,
we need to consider bremsstrahlung subprocesses, $GN\to 2GX$, or
$qN\to qGX$. The Born approximation for this processes includes
three graphs (interactions with the initial and two final partons).
The cross section of this reaction can be also expressed in terms of
the dipole approach \cite{kst2,kst1}, and can be eikonalized on a
nuclear target provided that the coherence length is sufficiently
long.

The condition of coherence is derived differently from the analysis
that lead to Eq.~(\ref{coh}). In this case $l_c$ is simply the
inverse longitudinal momentum transfer, \beq l_c=\frac{2E}{M^2},
\label{g-2g} \eeq where $E$ is the initial parton energy, and $M$ is
the invariant mass of the two final partons, which reads, \beq
M^2=\frac{p_T^2}{\alpha(1-\alpha)}. \label{inv-mass} \eeq Here $p_T$
is the relative transverse momentum of the final partons, $\alpha$
is the fractional light-cone momentum of one of the final partons,
and parton masses are neglected. Since gluon radiation is dominated
by small values of $\alpha\ll1$, Eqs.~(\ref{g-2g}) and
(\ref{inv-mass}) lead to the same expression for the coherence
length Eq.~(\ref{coh}), where $E_q$ should be replaced by the energy
of the parton detected in the final state.

In the long coherence length (LCL) regime, the transverse momentum spectra of gluon
bremsstrahlung for a high energy gluon interacting with a nucleon N
(nucleus A) target including the nonperturbative interactions of the
radiated gluon reads \cite{kst2,kst1},
\beqn
\frac{d\sigma(GN(A)\to G_1G_2X)}{d^2\vec{p}_T\,d^2\vec{b}} &=& \frac{1}{2(2\pi)^2}
\int d^2r_1d^2r_2
e^{i\vec p_T(\vec r_1-\vec r_2)}\overline{\Psi_{GG}^*(\vec r_1,\alpha )\Psi_{GG}(\vec r_2,\alpha )} \nonumber\\
&\times&
\im\Bigl[
f^{N(A)}_{3G}(\vec b,\vec r_{1},x)+
f^{N(A)}_{3G}(\vec b,\vec r_{2},x)
- f^{N(A)}_{3G}(\vec b,(\vec{r}_{1}-
\vec{r}_{2}), x)\Bigr],
\label{80}
 \eeqn
where $\alpha=P_{+}(G_1)/P_{+}(G)$ denotes the light-cone momentum
fractional of the radiated gluon.  The partial amplitude
$f^{N}_{3G}$ can be given in terms of the $q\bar{q}$ dipole
amplitude,
\beq
 \text{Im}f^{N}_{3G}(\vec{b},\vec{r},x)=\frac{9}{8}\{ \text{Im}f^{N}_{q\bar{q}}
 (\vec{b},\vec{r},x)+\text{Im}f^{N}_{q\bar{q}}(\vec{b},\alpha\vec{r},x)
+\text{Im}f^{N}_{q\bar{q}}(\vec{b},(1-\alpha)\vec{r},x)\},
\label{3g-1} \eeq where the factor $9/8$ is the ratio of Casimir
factors. Here the vectors $\vec r$, $\alpha \vec r$ and
$(1-\alpha)\vec r$ denote the two gluon transverse separations $\vec
r (G_1)-\vec r (G_2)$, $\vec r (G)-\vec r (G_2)$ and $\vec r
(G)-\vec r (G_1)$, respectively. Notice that the parameter $\beta$
that is present in the dipole saturation model of Eq.~(\ref{340})
corresponds here to the fraction of the total $3G$ momentum carried
by the $2G=G-G_2$ system, and is related to the light-cone momentum
fractional of the radiated gluon $\alpha$ by
 \beq
 \beta=1-\frac{\alpha}{2}.
\eeq

We still have to specify the light-cone distribution function
($\Psi_{GG}$) for $GG$ Fock component fluctuations of the incoming
gluon, which includes nonperturbative interactions of these gluons.
This is given by, \beqn \Psi_{GG}(\vec r,\alpha) &=&
\frac{\sqrt{8\alpha_s}}{\pi\,r^2}\,
\exp\left[-\frac{r^2}{2\,r_0^2}\right]\, \Bigl[\alpha(\vec
e_1^{\,*}\cdot\vec e)(\vec e_2^{\,*}\cdot\vec r) + (1-\alpha)(\vec
e_2^{\,*}\cdot\vec e)(\vec e_1^{\,*}\cdot\vec r) -
\alpha(1-\alpha)(\vec e_1^{\,*}\cdot\vec e_2^{\,*})(\vec e\cdot\vec
r) \Bigr],\ \label{100}
 \eeqn
where $r_0=0.3\fm$ is the parameter characterizing the strength of
the nonperturbative interaction, and which has been fitted to data
on diffractive $pp$ scattering \cite{kst2}. In Eq.~(\ref{80}) the
product of the wave functions is averaged over the initial gluon
polarization, $\vec e$, and summed over the final ones, $\vec
e_{1,2}$.

We consider here a case relevant for high energy gluon radiation
with $\alpha\to 0$. The azimuthal asymmetry of gluons $v_2^g$ coming
from a gluon-nucleon collision can be defined in a similar way to
the $v_2^q$ given in Eq.~(\ref{v2-1}), although replacing the
particle spectra with the one in Eq.~(\ref{80}).  After some algebra
one obtains, \beqn v_{2}^{gN}(p_{T},b)&=& \frac{\int_{-\pi}^{\pi}
d\phi~ \cos(2\phi) \frac{d\sigma(GN\to
G_1G_2X)}{d^2\vec{p}_T\,d^2\vec{b}} }
{\int_{-\pi}^{\pi} d\phi~ \frac{d\sigma(GN\to G_1G_2X)}{d^2\vec{p}_T\,d^2\vec{b}}}, \nonumber\\
&=&\frac{ \int_0^{\infty}dr \int_{-\pi}^{\pi}d\delta~ \cos(2\delta) \text{Im}f^{N}_{3G}(\vec{b},\vec{r})
\Biggl\{ \frac{2\pi}{p_T}\left(1-e^{-p_T^2r_0^2/2}\right) \left(J_1(p_T r)-J_3(p_T r)\right)
e^{\frac{-r^2}{2r_0^2}} +J_2(p_T r)e^{\frac{-r^2}{4r_0^2}} f(r,\delta)  \Biggl\}     }
{ \int_0^{\infty}dr \int_{-\pi}^{\pi} d\delta~ \text{Im}f^{N}_{3G}(\vec{b},\vec{r})
\Biggl\{ \frac{4\pi}{p_T}\left(1-e^{-p_T^2r_0^2/2}\right)J_1(p_T r) e^{\frac{-r^2}{2r_0^2}}-
J_0(p_T r)e^{\frac{-r^2}{4r_0^2}}f(r,\delta)
 \Biggl\}},
 \label{vn1}\nonumber\\
\eeqn 
where the function $f(r,\delta)$ is defined as
\begin{eqnarray}
f(r,\delta)&=& \int_{0}^{\infty} d\Delta \int_{-\pi}^{+\pi} d \theta \frac{(\Delta^2-r^2)
\Delta r}{(\Delta^2+r^2)^2-4(\Delta r\cos(\delta-\theta))^2}
e^{-\frac{\Delta^2}{4r_0^2}}. \label{not}\
\end{eqnarray}

In the case of a nuclear target Eq.~(\ref{80}) still holds, but the
dipole amplitude on a nucleon target $f^{N}_{3G}$ should be replaced
with the one on a nuclear target $f^{A}_{3G}$. The partial elastic
amplitude $f^{A}_{3G}$ for a colorless three-gluon system colliding
with a nucleus can be written in terms of the partial amplitude
$f^{N}_{3G}$ of a three-gluon system colliding with a proton at
impact parameter $b$, \beq \text{Im}f^{A}_{3G}(\vec{b},\vec{r},x)= 2
\Biggl\{1-\exp[-\int d^{2}\vec{s}~
\text{Im}f^{N}_{3G}(\vec{s},\vec{r},x)T_{A}(\vec{b}+\vec{s})]\Biggl\},
\label{eik2} \eeq where the $3G$ amplitude $f^{N}_{3G}$ is related
to the $q\bar{q}$ dipole amplitude via Eq.~(\ref{3g-1}). In a very
similar fashion, one can obtain the gluons $v_2$ in a gluon-nucleus
collision (see Appendix A for a derivation), \beqn
v_{2}^{gA}(p_{T},b)&=& \frac{\int_{-\pi}^{\pi} d\phi~ \cos(2\phi)
\frac{d\sigma(GA\to G_1G_2X)}{d^2p_T\,d^2b}  } {\int_{-\pi}^{\pi}
d\phi~ \frac{d\sigma(GA\to G_1G_2X)}{d^2p_T\,d^2b}  }=\frac{
\int_0^{\infty}dr \int_{-\pi}^{\pi}d\delta~ \cos(2\delta)
\Psi_N(p_T, r, b, \delta)} { \int_0^{\infty}dr
\int_{-\pi}^{\pi}d\delta~ \Psi_D(p_T, r, b, \delta)+2\pi g(p_T)},
\label{vgn}\ \eeqn where we defined
\beqn \Psi_N(p_T, r, b,
\delta)&=&-\frac{(2\pi)^2}{p_T}\left(1-e^{-p_T^2r_0^2/2}\right)
\left(J_1(p_T r)-J_3(p_T r)\right)
e^{\frac{-r^2}{2r_0^2}-\mathcal{I}_G(b, r,\delta)}
-2\pi J_2(p_T r)e^{\frac{-r^2}{4r_0^2}-\mathcal{I}_G(b, r,\delta)}f(r,\delta),\nonumber\\
\Psi_D(p_T, r, b, \delta)&=& -\frac{2(2\pi)^2}{p_T}\left(1-e^{-p_T^2r_0^2/2}\right)J_1(p_T r) e^{\frac{-r^2}{2r_0^2}-\mathcal{I}_G(b, r,\delta)}
+2\pi J_0(p_T r)e^{\frac{-r^2}{4r_0^2}-\mathcal{I}_G(b, r,\delta)}f(r,\delta)\nonumber\\
g(p_T)&=&\frac{(2\pi)^2}{p_T^2}\left(1-e^{-p_T^2r_0^2/2}\right)^2,\
\eeqn with the notation,
\begin{eqnarray}
\mathcal{I}_G(b, r,\delta)&=&
\frac{9}{4}\int d^{2}\vec{s}~\text{Im}f^{N}_{q\bar{q}}(\vec{s},\vec{r})T_{A}(\vec{b}+\vec{s}),\label{not1}\
\end{eqnarray}

The remaining integrals in
Eqs.~(\ref{vn1},\ref{not},\ref{vgn},\ref{not1}) can be performed
only numerically. From Eqs.~(\ref{vn1},\ref{vgn}) one can observe
that similar to Eq.~(\ref{sv2}), the angle dependence $\phi$ between
the impact parameter $\vec{b}$ and the transverse momentum of the
projectile gluon $\vec{p}_{T}$ was replaced by the angle $\delta$
between $\vec{b}$ and dipole vector $\vec r$. As a consequence the
azimuthal asymmetry of the radiated gluon is directly related to the
orientation of the color dipole.

\section{Pions azimuthal asymmetry in $pp$ and $pA$ collisions}

The invariant cross section for hadron production in $pp$ collisions
can be described, in the pQCD-improved parton model based on
factorization \cite{fields-feynman,own,wang}, by the expression
\begin{eqnarray}
\frac{d \sigma^{pp\to h+X}}{dy d^2 p_T} &=&  \sum_{ijkl}\int dx_i dx_j d^2 k_{iT}d^2 k_{jT}
f_{i/p}(x_i,Q^2)\mathcal{G}_p(k_{iT},Q^2) f_{j/p}(x_j,Q^2)\mathcal{G}_p(k_{jT},Q^2)
K\,\frac{d
\sigma}{d \hat{t}}(ij \to kl) \, \frac{D_{h/k}(z_k,Q^2)}{\pi z_k}, \label{fact}\nonumber\\
\end{eqnarray}
where we sum over different species of participating
partons, and $f_{i/p}(x_i,Q^2)$, $f_{j/p}(x_j,Q^2)$ are the parton
distribution functions (PDF) of the colliding protons, which depend
on the light-cone momentum fractions $x_i$, $x_j$ and the hard scale
$Q$. The function $D_{h/k}(z_k,Q^2)$ is the fragmentation function
of parton $k$ to the final hadron $h$ with a momentum fraction
$z_k$. The cross section $\frac{d \sigma}{d \hat{t}}(ij \to kl)$ of
the hard process, which is a function of Mandelstam variables, can
be calculated perturbatively.

The higher order perturbative corrections are taken into account via
a $K$-factor, and the primordial momentum distributions $
\mathcal{G}_p(k_{T},Q^2)$, which are assumed to have the form
\begin{equation}
\mathcal{G}_p(k_{T},Q^2) = \frac{\textrm{exp}(-k_{T}^2/\langle k^2_T
\rangle)}{\pi \langle k^2_T \rangle },\label{kt}
\end{equation}
with the mean values $\langle k^2_T \rangle$, are taken to be
independent of $Q^2$ for the sake of simplicity.

\subsection{Short coherence length regime}

 An important QCD prediction is the $p_T$ power dependence of jet production, confirmed by data.
Due to this property the multiple interactions of a large $p_T$
produced parton do not share equally the total transferred momentum
$p_T$ (like it would be if the $p_T$ dependence of each collision
were Gaussian). In fact, there is one collision with a large
transverse momentum, close to $p_T$, while the others are mainly
soft interactions with small transferred momenta. Equation
(\ref{main}) fully includes this dynamics, although it is really
valid only for coherent multiple rescatterings, when multiple
interaction amplitudes, rather than cross sections, are convoluted.
What happens if one decreases the energy, or increases $p_T$, and
eventually gets into a regime of short coherence length
(SCL)(\ref{coh})? According to the specific QCD dynamics  described
above, only the single high-$p_T$ collision becomes incoherent,
while the other multiple soft collisions remain coherent. Thus, we
arrive at a three step picture: soft multiple coherent interactions
slightly increasing the transverse momentum of the parton, followed
by a hard incoherent parton-nucleon collision with high $p_T$, and
eventually multiple soft final state interactions of the produced
parton leading to an additional broadening. A proper tool for
calculations of soft multiple broadening is the dipole approach
\cite{boris0}, which allows to use the phenomenology of soft
interactions. At the same time, for the hard parton-nucleon
collision we rely on the factorization based parton model, since at
large $p_T$ Bjorken $x$ of the target is too large for the dipole
technique to be valid.

Thus, the inclusive cross section of $pA\to hX$ can be written as
\begin{eqnarray}
\frac{d \sigma^{pA\to h+X}}{dy d^2 \vec p_T d^2 \vec b} &=& 
T_A(b)\sum_{ijkl}\int dx_i dx_j d^2 k_{iT}d^2 k_{jT}
f_{i/p}(x_i,Q^2)\,\tilde{\mathcal{G}}_p(k_{iT},Q^2) 
\tilde f_{j/N}(x_j,Q^2) 
\mathcal{G}_p(k_{jT},Q^2)\nonumber\\
&&\hspace{1.5cm}\times K\, \frac{d \sigma}{d \hat{t}}(ij \to kl) \frac{D_{h/k}(z_k,Q^2)}{\pi z_k}.
\label{fact1}\
\end{eqnarray}
The parton distribution function of a bound
nucleon, $\tilde f_{j/N}(x_j,Q^2)$, is known to be modified by the
nuclear environment, a phenomenon called EMC effect \cite{emc}.
Notice that shadowing effects should not be considered, since we
need the PDF of a single bound nucleon. Moreover, in the SCL regime
the coherence length is too short for any shadowing effects to
appear. At large $x_j$ isotopic effect may be important for the
target nucleon PDF, therefore we average over the nucleus, \beq
\tilde f_{j/N}(x,Q^2)=\frac{Z}{A}\tilde f_{j/p}(x,Q^2)+
\left(1-\frac{Z}{A}\right)\tilde f_{j/n}(x,Q^2), \label{npdf} \eeq
with atomic and charge numbers $A$ and $Z$ respectively.

Initial/final state broadening of the projectile/ejectile partons is
effectively taken into account via a modification of the primordial
transverse momentum distribution, $\mathcal{G}_p(k_{iT})\Rightarrow
\tilde{\mathcal{G}}_p(k_{iT})$, where 
\beq
\tilde{\mathcal{G}}_p(k_{iT})= \frac{d
N^{iA\to iX}(b)}{d^{2}k_{iT}}. \label{broad} 
\eeq 
The $k_T$-distribution $dN^{iA\to iX}(b)/d^{2}k_{T}$, normalized to
unity, is calculated using Eq.~(\ref{main}) and the KST
parametrization of the dipole cross section (with a Casimir factor 9/4
for gluons). Notice that in the SCL regime under consideration the
process of broadening is dominated by soft coherent multiple
interaction which have no relation to the hard scale $Q^2$ imposed by
the high-$p_T$ process occurring incoherently. Therefore in this case
the density matrix Eq.~(\ref{kkt}) should have a rather small mean
value of the primordial quark momentum $\la \tilde k_T^2\ra\sim
\Lambda_{QCD}^2$. To simplify the calculations we assume that the
initial and final partons are the same, so the total nuclear
thickness $T_A(b)$ contributes to broadening.

\subsection{Midrapidities at high energies}

At high energies and midrapidities the parton fractional momenta in
the beam and target are small, $x_1\sim x_2\sim 2p_T/\sqrt{s}\ll1$,
so hadron production is dominated by fragmentation of radiated
gluons, and we can rely on the results of Section~V. The cross
section of hadron production in $pp$ collisions at impact parameter
$\vec b$ is then given by a convolution of the distribution function
of the projectile gluon inside the proton with the gluon radiation
cross section coming from $GN$ collisions and also with the
fragmentation function, 
\beq
\frac{d \sigma^{pp\to h+X}}{dy d^2 \vec p_T d^2 \vec b}= \int dx_g f_{G/p}(x_g,Q^2) 
\frac{d\sigma(Gp\to G_1G_2X)}{d^2k_{gT}\,d^2b} \frac{D_{h/G_2}(z,Q^2)}{z^2}.
\label{pp1}
\eeq
To simplify the calculations we assume here that the projectile
gluon has the same impact parameter relative to the target as the
beam proton. At midrapidities we have $x_g=2k_{gT}/\sqrt{s}$ and we
take $Q^2=k^2_{gT}$. The cross section of gluon radiation in the
above expression can be obtained by the master Eq.~(\ref{80}).  This
cross section reproduces well the measured pion cross section in
$pp$ collisions \cite{kst1,boris2}. In the LCL regime, a high $p_T$
parton propagating through the nucleus is freed by the multiple
coherent interactions, and the Cronin effect may be conceived as
color-filtering. In this regime the cross section of hadron
production in $pA$ collisions has the form, 
\beq
\frac{d \sigma^{pA\to h+X}}{dy d^2 \vec p_T d^2 \vec b}= \int dx_g f_{G/p}(x_g,Q^2) \frac{d\sigma(GA\to G_1G_2X)}{d^2k_{gT}\,d^2b} \frac{D_{h/G_2}(z,Q^2)}{z^2},
\label{pp2}
\eeq
where the cross section of gluon
radiation in $GA$ collisions can be obtained from
Eqs.~(\ref{80},\ref{eik2}). In Eqs.~(\ref{pp1},\ref{pp2}), we will
use the same PDFs and the fragmentation functions which are employed
in Eq.~(\ref{fact}).

\subsection{Azimuthal asymmetry of produced hadrons}

Notice that the asymptotic expressions (\ref{pp1}, \ref{pp2}),
supplemented with the gluon radiation cross-section given in
Eq.~(\ref{80}) at $\alpha\ll1$, are only reliable at very long
coherence lengths, which is certainly the case at LHC energies. At
RHIC energies, for hadrons produced at midrapidities with moderate
$p_T$, we are in the transition region between the regimes of long
and short coherence lengths. However, for peripheral collisions
where $v_2$ is not zero, we are in the LCL regime.

The azimuthal asymmetry of hadrons produced in $pp$ and $pA$
collisions at impact parameter $b$ is computed as a second order
Fourier coefficients in a Fourier expansion of the azimuthal
dependence of the transverse momentum spectra of the inclusive
hadron production, Eqs.~(\ref{fact1},\ref{pp1},\ref{pp2}), around
the beam direction,
\begin{equation}
v_{2}^{\pi}(p_{T},b)= \frac{\int_{-\pi}^{\pi} d\phi \cos(2\phi) \frac{d \sigma^{pp(A)\to h+X}}
{dy d^2 \vec p_T d^2 \vec b} }
{\int_{-\pi}^{\pi} d\phi \frac{d \sigma^{pp(A)\to h+X}}{dy d^2 \vec p_T d^2 \vec b}}. \label{v2-pi}
\end{equation}
In the SCL regime, the main element of the hadron production
Eq.~(\ref{fact1}), which leads to an azimuthal asymmetry, is the
angular dependence of the broadened projectile partons via
Eq.~(\ref{mainv}).  At the same time, in the LCL regime, the angle
dependence of the gluon radiation cross section Eq.~(\ref{80})
induces via Eqs.~(\ref{pp1},\ref{pp2}) an azimuthal asymmetry for
the produced hadrons. In both cases, the introduction of the
color-dipole orientation is the key ingredient.

\begin{figure}
\includegraphics[height=.30\textheight]{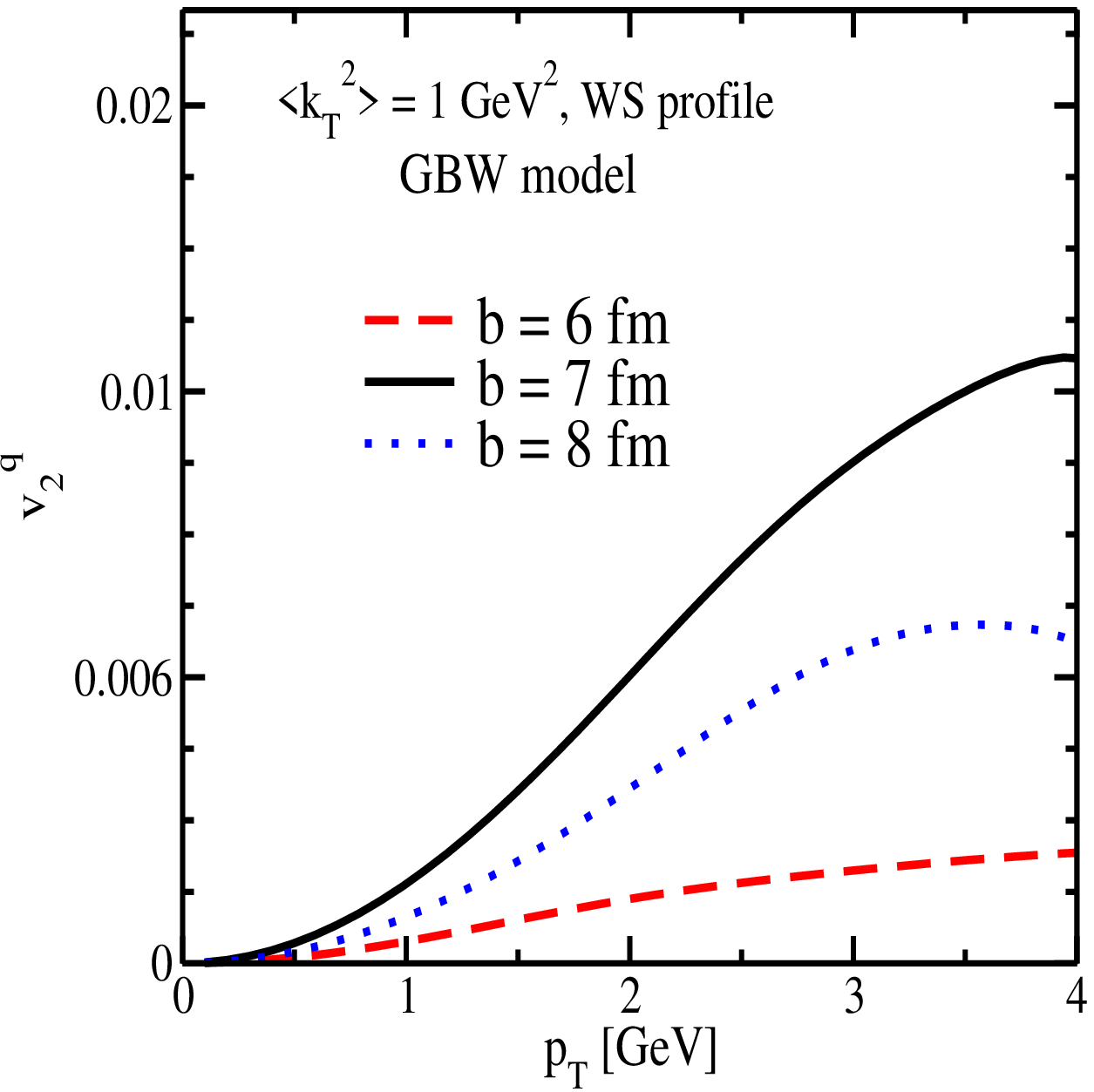}
 \includegraphics[height=.30\textheight]{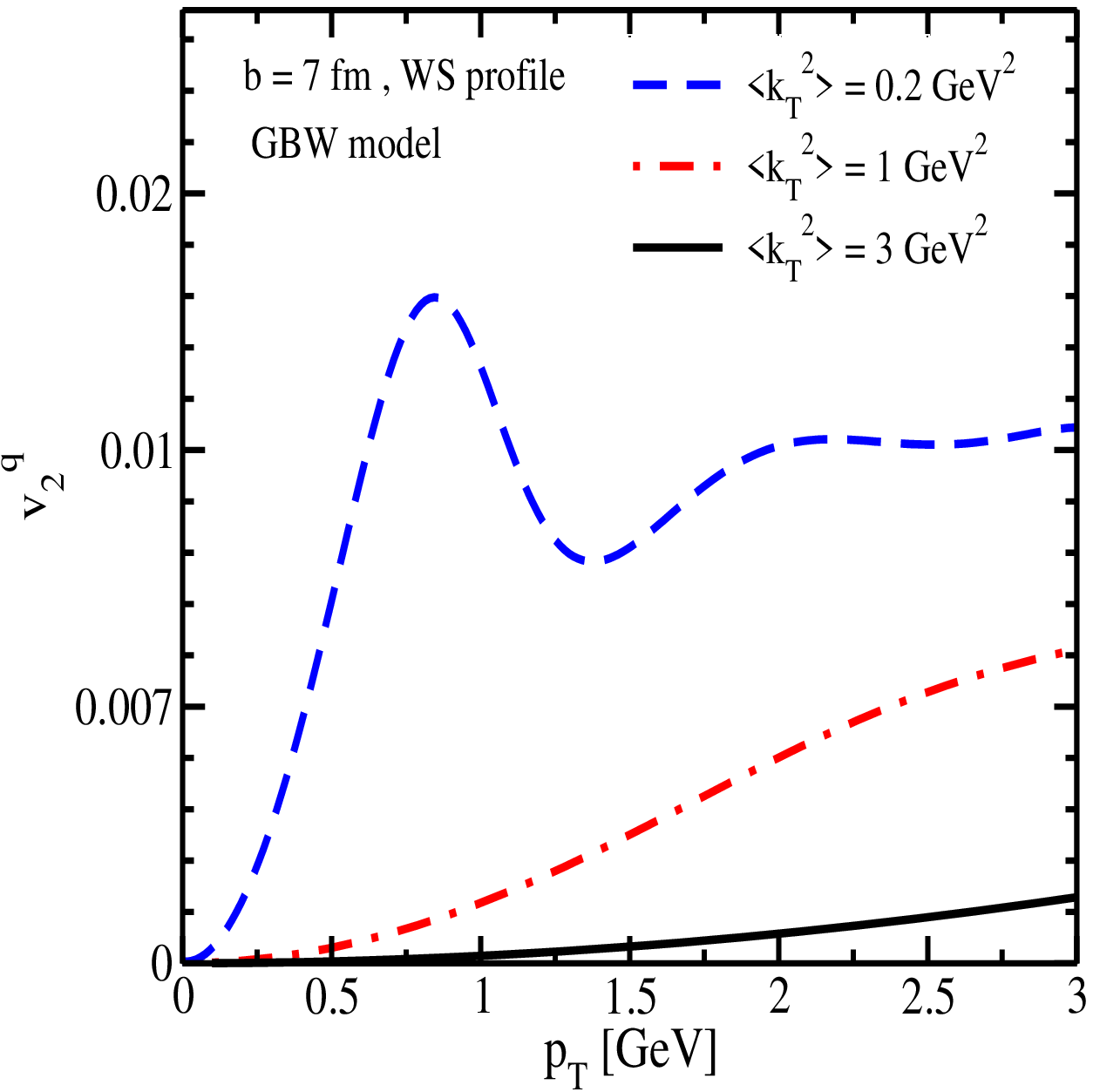}
\caption{Right: The
 azimuthal anisotropy of quarks from quark-nucleus
 collisions, obtained from Eq.~(\ref{sv2}), for various primordial
 transverse momentum squared $\langle k_{T}^{2}\rangle$ at a fixed
 impact parameter $b=7$. Left: The
 azimuthal anisotropy of quarks from quark-nucleus
 collisions at various impact parameters, with a fixed primordial
 transverse momentum squared $\langle
 k_{T}^{2}\rangle=1~\text{GeV}^2$.  All curves are obtained for the RHIC energy $\sqrt{s}= 200$ GeV,
 within the GBW dipole model with
Woods-Saxon (WS) nuclear profile. \label{fig0} }
\end{figure}

\section{Numerical results}
First, we show the $v_2$ of quarks and
gluons in parton-nucleus collisions introduced in sections IV and
V. The only external input here is the nuclear profile. We take first the
Woods-Saxon (WS) profile, with a nuclear radius $R_{A}=6.5$ fm and a
surface thickness $\xi=0.54$ fm. In our approach, the profile of
nuclear density at the edge is a very important input, since the
elliptic asymmetry stems from the rapid change of nuclear density at
the edge.  In order to show this more clearly, we also show the
results for the hard sphere (HS) nuclear profile with a constant
density distribution, $\rho_{A}=\rho_{0}\Theta(R_{A}-r)$, with the
same nuclear radius as was taken for the WS profile.

In Fig.~(\ref{fig0}), we show $v_2$ of quarks at various impact
parameters, within the saturation model I, and at the RHIC energy. A
smaller primordial transverse momentum $\langle k_{T}^{2}\rangle$
leads to a bigger broadening and more multi-scatterings,
consequently the azimuthal asymmetry will be also bigger.

\begin{figure}
 \includegraphics[height=.30\textheight]{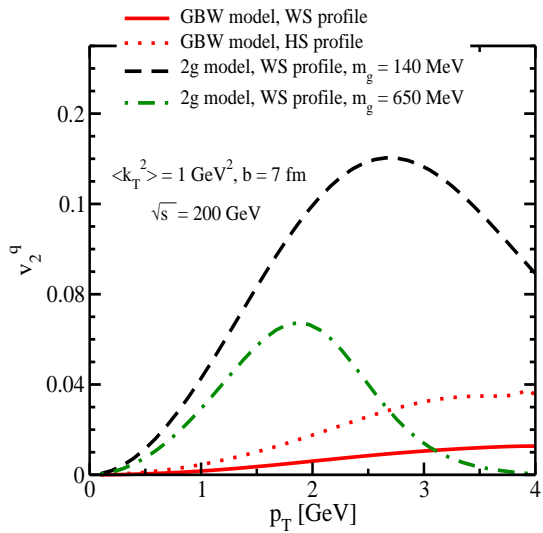}
\caption{ The azimuthal anisotropy of quarks coming from
quark-nucleus scatterings,
 for both the Woods-Saxon (WS) and hard sphere (HS) profiles, within
 the GBW dipole model and the improved Born approximation model,
 at the RHIC energy $\sqrt{s}= 200$ GeV. For the 2g model, we show the $v^{q}_2$ for two
 different effective gluon masses $m_g=140$ and $650$ MeV . \label{fig1}}
\end{figure}

The main source of azimuthal asymmetry in the amplitude
(\ref{eik},\ref{eik2}) is the interplay between multiple rescatering
and the shape of the physical system.  The key function which
describes the effect of multiple interactions is the eikonal
exponential and the color dipole amplitude. The information about
the shape of the system is incorporated through a convolution of the
impact parameter dependent partial elastic amplitude and the nuclear
thickness function. The initial space-time asymmetry gets then
translated into a momentum space anisotropy by the double Fourier
transform in Eqs.~(\ref{main}, \ref{80}).  For more central
collisions, the correlation between nuclear profile and dipole
orientation is minimal. In fact if the nuclear thickness was
constant, then the convolution between the nuclear profile and the
dipole orientation would become trivial and there would be no
azimuthal asymmetry. Therefore, the main source of azimuthal
anisotropy is not present for central collisions where the nuclear
density has only small variation. This can be seen in
Figs.~(\ref{fig0},\ref{fig1}), where a pronounced elliptic
anisotropy is observed for collisions with impact parameters close
to the nuclear radius $R_{A}$, where the nuclear profile undergoes
rapid changes. In Fig.~(\ref{fig1}), for comparison, we show the
$v_2^q$ coming from the WS and the HS nuclear profile, within the
GBW model. The $v_2^q$ of quark with the HS nuclear profile can be
twice bigger than the one with WS nuclear profile. This indicates
that the nuclear density profile is an important input.

In Fig.~(\ref{fig1}), we show the $v_2^q$ of quarks in $pA$
collisions within the GBW model and the improved Born approximation,
the 2g model. For the 2g model, we show $v_2^q$ for two different
effective gluon masses $m_g=140$ and $650$ MeV.   There are growing
evidences suggesting that the effective gluons mass should be bigger
than the confinement scale. For instance, a small gluon correlation
radius $0.35$ fm is a result of lattice calculations \cite{lat}, and
it is also predicted by the instanton model \cite{ins}. Experimental
data suggests an enhanced intrinsic motion of gluons in light hadrons
compared to the inverse hadronic radius \cite{kst1}. And also the
smallness of the triple-Pomeron coupling can be only explained by a
enhanced gluon interactions due to nonperturbative effects
\cite{spot}.

It is obvious from Fig.~(\ref{fig1}) that the azimuthal asymmetry of
quarks is very sensitive to the effective gluon mass and the higher
order corrections. The azimuthal asymmetry within the saturation
model which includes higher order radiation corrections is
significantly smaller than the one for the 2g model. Therefore, the
inclusions of the higher-order radiation corrections or taking a
bigger effective gluon mass reduces the azimuthal asymmetry.  A
quark propagating through a nucleus interacts by gluon exchange with
different nucleons located at different azimuthal angles relative to
the quark trajectory, and their contributions to $v_2$ tends to
cancel each other, reducing the azimuthal symmetry.

\begin{figure}
\includegraphics[height=.30\textheight]{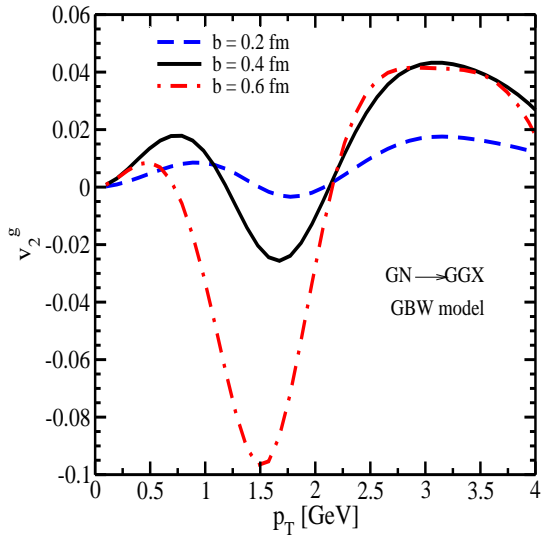}\\
\caption{The
 azimuthal anisotropy of gluons from gluon-nucleon
 collisions, given by Eq.~(\ref{vn1}), at various impact parameters, within the GBW dipole
 model, and with the WS nuclear profile. \label{fig2} }
\end{figure}

\begin{figure}
\includegraphics[height=.30\textheight]{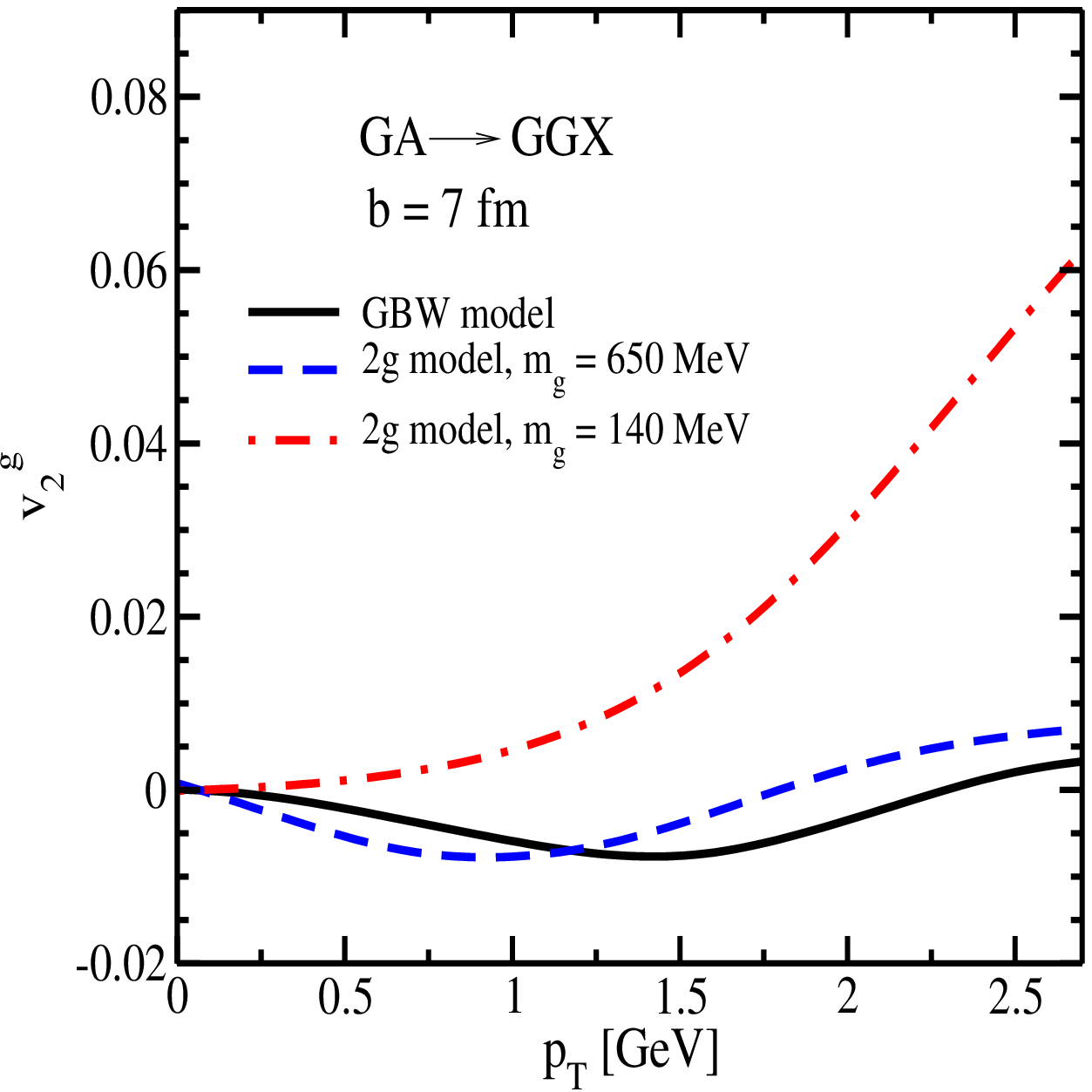}
\includegraphics[height=.30\textheight]{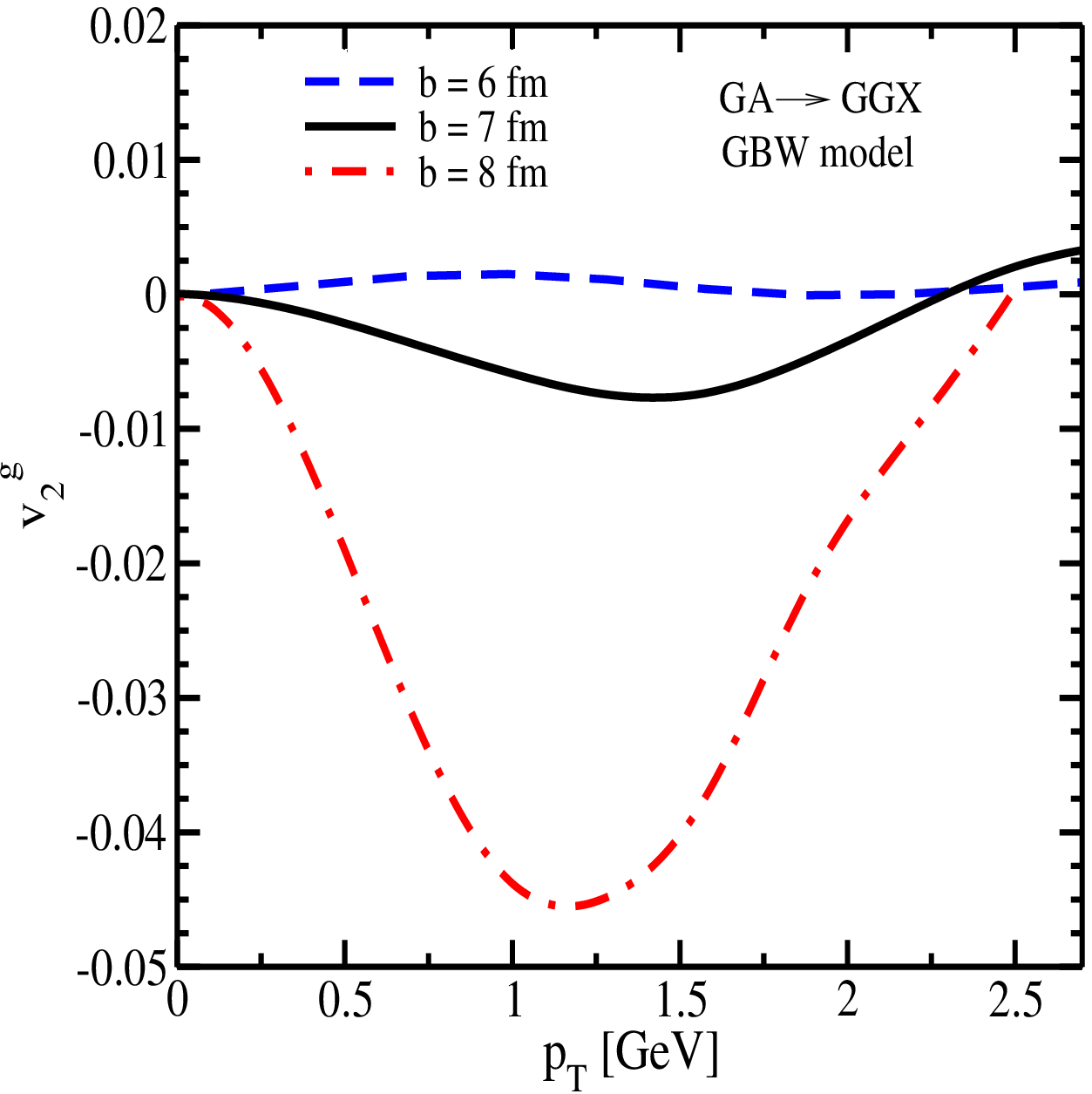}
\caption{Right: The
 azimuthal anisotropy of gluons from gluon-nucleus
 collisions, given in Eq.~(\ref{vgn}), at various impact parameters, within the GBW dipole model. Left:
 $v_2^g$ at a fixed impact parameter $b=7$ fm, within various dipole models. We used for all
 curves the WS nuclear profile. \label{fig3} }
\end{figure}

In Figs.~(\ref{fig2},\ref{fig3}), we show the azimuthal asymmetry of
gluon radiation for gluon-nucleon and gluon-nucleus collisions,
given in Eqs.~(\ref{vn1},\ref{vgn}) at various impact parameters.
The azimuthal asymmetry of gluons in both $GN$ and $GA$ has a rather
similar trend; it can be negative at small $p_T$ for peripheral collisions and becomes positive at
higher $p_T$. Again, for more peripheral collisions where the color
dipole orientation becomes more important, the azimuthal asymmetry
is bigger. In Fig.~(\ref{fig3}), on the left panel, we also show the
gluonic $v_2$ in $GA$ collisions at a fixed impact parameter $b=7$
fm, within various color dipole models. Notice that for the case of
gluon bremsstrahlung we have already included some nonperturbative
effects of gluon interactions through the $GG$ distribution function
Eq.~(\ref{100}), via the parameter $r_0=0.3$ fm, which simulates the
strength of the non-perturbative gluon interactions contained in the
diffractive $pp$ scattering data. Therefore, the azimuthal
asymmetries of gluons $v_2^g$ within the saturation dipole model and
the improved Born approximation with effective gluon mass $m_g=650 $
MeV are rather similar.

Next, we calculate the azimuthal asymmetry of produced pions in $pp$
and $pA$ collisions, within two different schemes, the standard pQCD
factorization Eqs.~(\ref{fact},\ref{fact1}) (short-coherence) and
light-cone factorization Eqs.~(\ref{pp1},\ref{pp2})
(long-coherence).

\begin{figure}[!t]
\includegraphics[height=.30\textheight]{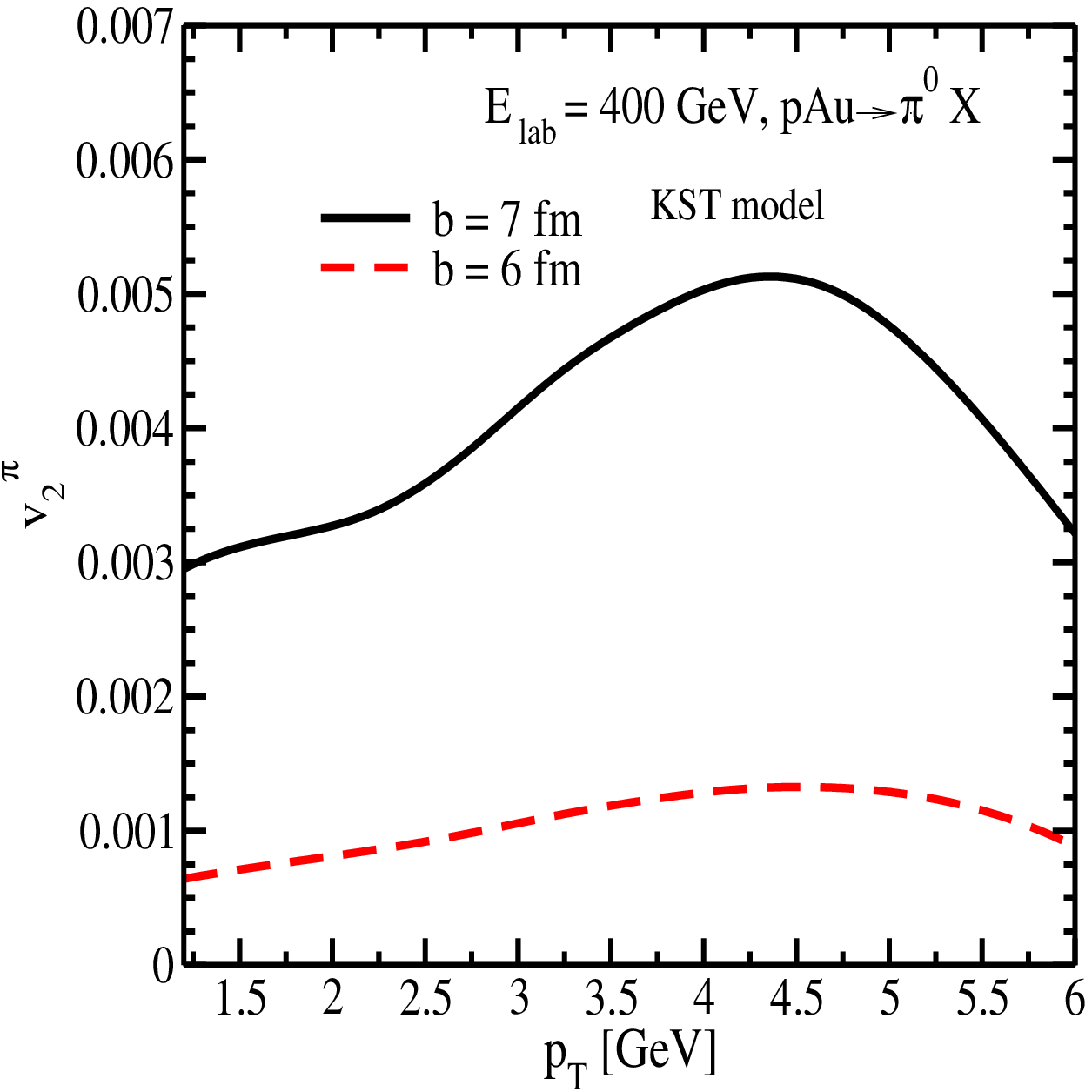}
 \includegraphics[height=.30\textheight]{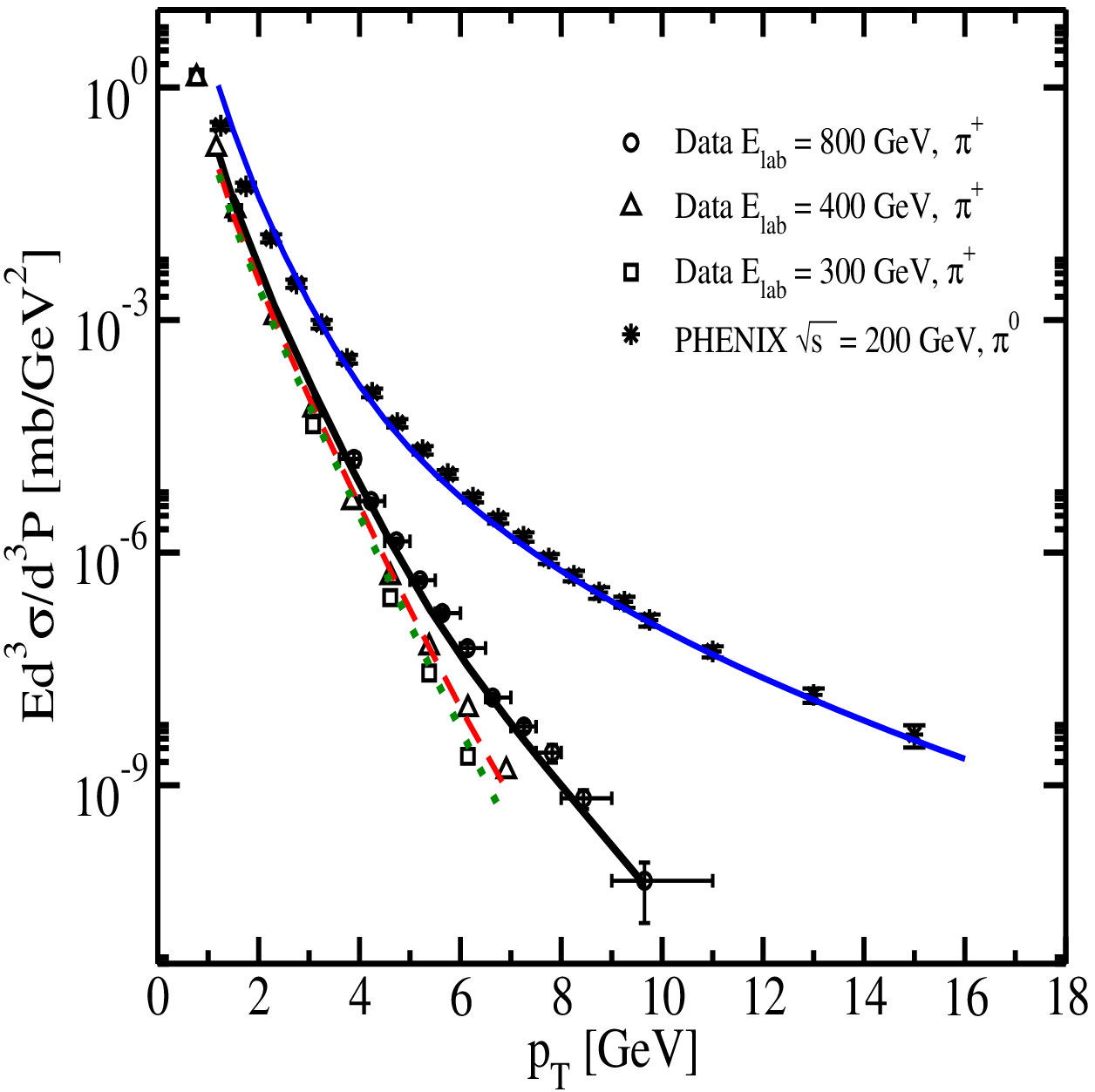}
 \caption{Right: the cross section of $p + p \to \pi^0 + X$ at RHIC
 for $\sqrt{s}= 200$ GeV, and that of $p + p \to \pi^+ + X$ at the SPS
 fixed target experiment for $E_{lab} = 300$, 400 and 800 GeV. Data
 for RHIC and SPS are from Ref.~\cite{rhic2006} and
 Refs.~\cite{sps79,sps89}, respectively. Left: The azimuthal
 anisotropy of the produced pions $v_2^{\pi}$ in $pA$ collisions, in the short-coherence regime, at
 energy $E_{lab} = 400$ GeV, for the Woods-Saxon (WS) nuclear profile, and within the KST dipole model.
\label{fig4}}
\end{figure}

We first concentrate on the short-coherence scheme. For the parton
distribution $f_{i/p}(x,Q^2)$ used in our calculations, we employ
the MRST (2006 NNLO) set~\cite{mrst2006}. For the fragmentation
function $D(z,Q^2)$, we use the parametrization given by
Kretzer~\cite{kretzer2000}, with NLO corrections.  In our
calculation we will take $\langle k^2_T \rangle = 3 \,
\textrm{GeV}^2 $. This value corresponds to an average transverse
momentum of the parton $k_T \approx 1.5$ GeV, which coincides with
the analysis of $k_T$ for photon and pion production given in
Refs.~\cite{ktc2,apanasevich98}. Notice also that different
value of $\langle k^2_T \rangle$ has been used in different
approaches \cite{ktc3}. For the scale $Q$ of the hard
process in Eqs.~(\ref{fact},\ref{fact1}), we choose $Q = p_T/(1.2
z_c)$. We take the $K$-factor $K\approx 2$, which gives a good
approximation of the NLO order contribution in the $p_T$ region of
interest. With this setup, we are able to describe pion production
data for $pp$ collisions from the SPS to RHIC energies within $35\%$
discrepancy, see Fig.~(\ref{fig4}) right panel.

In this way, all free phenomenological parameters for hadron
production in $pA$ collisions, Eq.~(\ref{fact1}), are already fixed
in reactions different from $pA$ collisions. We recall that the
azimuthal asymmetry originates here from the rapid change of the
nuclear density, and therefore only the peripheral tail of the
nuclear profile contributes to the azimuthal asymmetry.  At the very
periphery, the nuclear parton distribution is unchanged compared to
a free nucleon. Nevertheless, we have numerically verified that
$v_2^{\pi}$ changes less than $20\%$ with various PDF
parametrizations for impact parameters bigger than $b>6$ fm. In
Fig.~(\ref{fig4}) we show the azimuthal asymmetry of the produced
pions in $pAu$ collision at midrapidity, for the fixed target energy
$E_{lab}=400$ GeV obtained in the SCL scheme, Eq.~(\ref{fact1}).

\begin{figure}[!t]
 \includegraphics[height=.30\textheight]{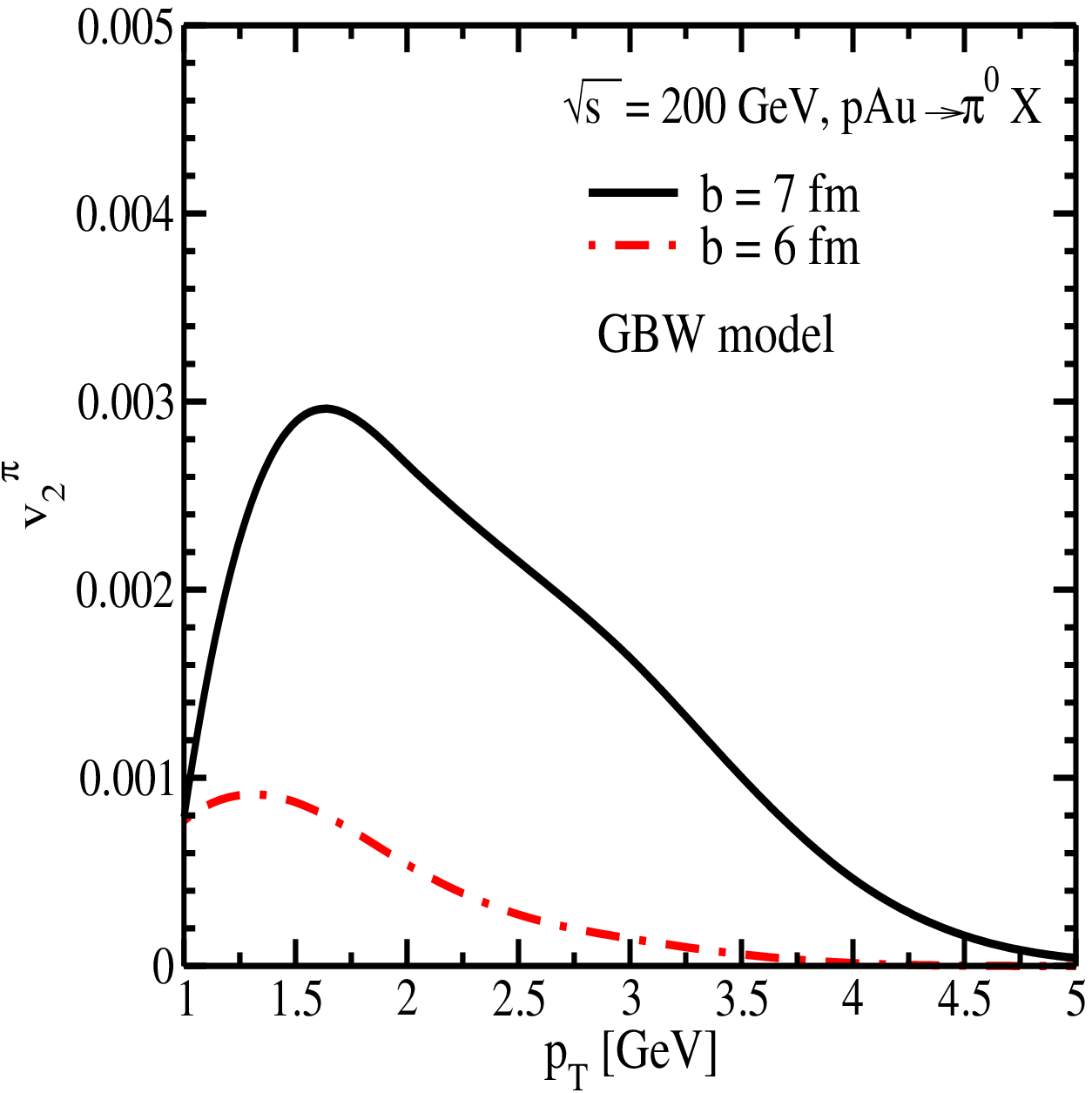}
 \includegraphics[height=.30\textheight]{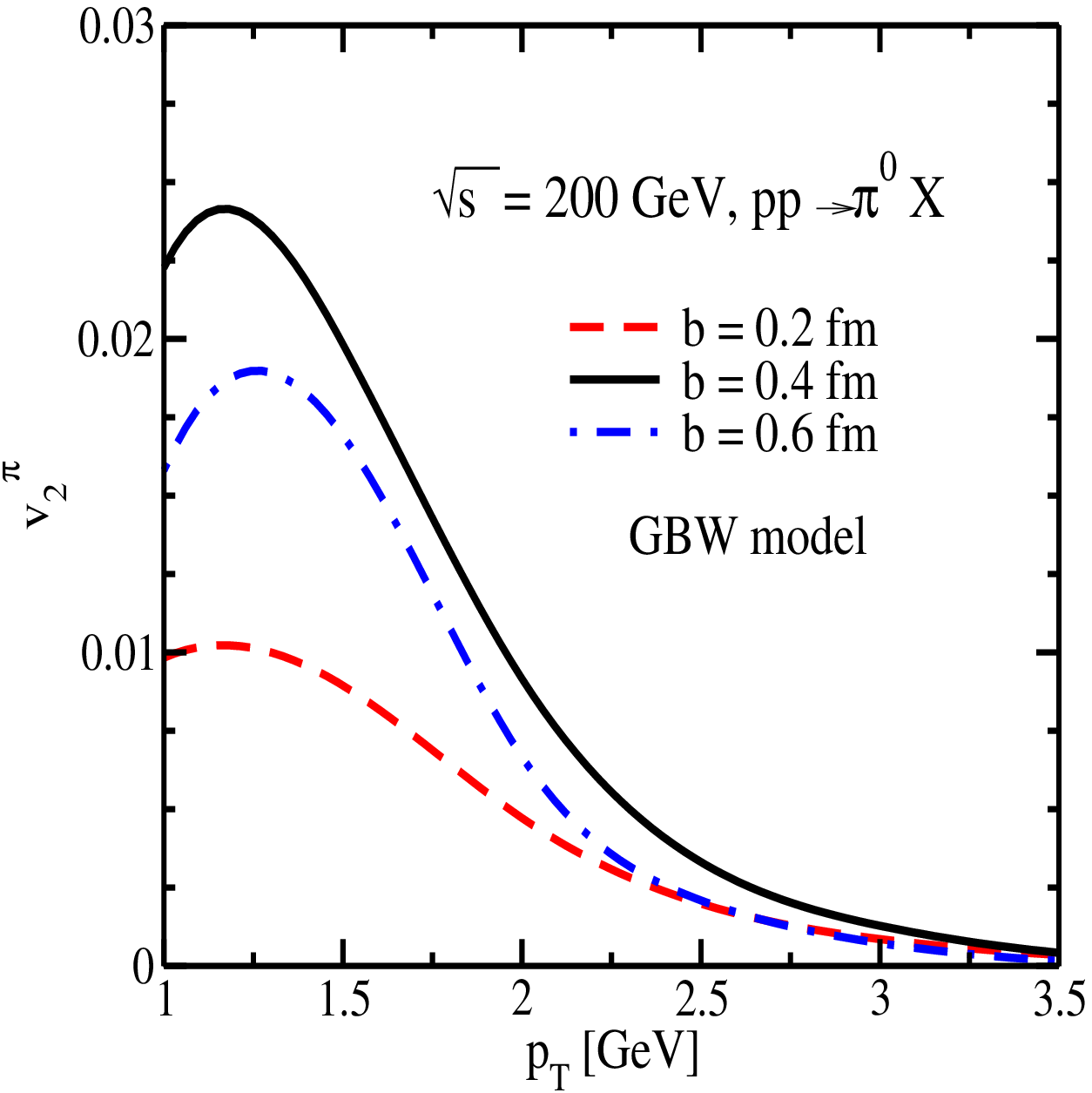}
\caption{Right: The
 azimuthal anisotropy of produced pions $v_2^{\pi}$ in $pp$ collision
 at the RHIC energy $\sqrt{s}= 200$ GeV, at various impact
 parameters. Left: The azimuthal anisotropy of produced pions in $pAu$
 collision with the WS nuclear profile at the RHIC energy. We used for
 all curves the GBW dipole model.\label{fig6}}
\end{figure}

In Fig.~(\ref{fig6}), right panel, we show the azimuthal asymmetry
of produced pions in $pp$ collisions at the RHIC energy $\sqrt{s}=
200$ GeV. A pronounced positive azimuthal asymmetry is observed, at
an impact parameter around half of the proton size $b\sim 0.5$ fm.
 The source of the azimuthal asymmetry
of pions in $pp$ collisions is the angle dependence of the gluon
bremsstrahlung cross section in Eqs.~(\ref{pp1}), via the
color-dipole orientation. The azimuthal asymmetry of the fragmented
pions in $pp$ collisions is sizable at a $p_T$ where the
color-dipole size becomes compatible with the impact parameter and
consequently the color-dipole orientation becomes important, see
Eq.~(\ref{340}).


At the LCL limit, pions are entirely produced by gluons. It may be
then puzzling that the pion's $v_2$ is positive while gluons have
negative $v_2$ in the same range of transverse momentum.
Nevertheless, one should notice that the gluons transverse momentum
$k_{gT}$ is related to the transverse momentum of the fragmented
pions via $k_{gT}=p_{T}/z$. Therefore, high $k_{gT}$ gluons with
positive $v_2^{g}$ are responsible for the produced pions at
moderate $p_T$, see Fig.~(\ref{fig2}).

In Fig.~(\ref{fig6}) left panel, we show the $v_2^{\pi}$ of the
produced pions in $pAu$ collisions, obtained in the LCL scheme.  The azimuthal asymmetry $v_2^{\pi}$
increases from zero at an impact parameter around nuclear radius to
a maximum value at an impact parameter around $b\sim 7 $ fm, and
then decreases again for a more peripheral collision.

\section{Summary and conclusions}
This paper is the first attempt to calculate the azimuthal asymmetry
$v_2$ of produced hadrons in $pp$ and $pA$ collisions.  We
generalized the hadron production scheme within the pQCD parton
model and the light-cone QCD-dipole formalism by the inclusion of
the color-dipole orientation. This is the key ingredient which leads
to the azimuthal asymmetry both in $pp$ and $pA$ collisions. We
introduced the color-dipole orientation into the improved Born
approximation and within the saturation model of Golec-Biernat and
W\"usthoff, which satisfies available DIS data.

In the short-coherence regime, we used the improved pQCD parton
model. The salient part of hadron production in $pA$ collisions at
the short-coherence regime is the broadening of the projectile
partons going through nucleus. This broadening is calculated here in
a free-parameter formalism which uses the color-dipole approach,
where the angle dependence of the broadened projectile partons via
the color-dipole orientation leads to the azimuthal asymmetry of the
produced hadron. In the long-coherence regime, we used the
light-cone dipole formalism in the rest frame of the target, which
is valid at small $x_2$. The main source of the azimuthal asymmetry
in the long-coherence regime originates from the angle dependence of
the radiated gluons cross section. Our results show that the
azimuthal asymmetry of pions in both $pp$ and $pA$ collisions is
rather small. This indicates that the contribution of the initial
state effects, which is present in cold nuclear matter in the
observed azimuthal asymmetry $v_2$ of the produced hadrons in $AA$
collisions at RHIC, is very small.

We have also systematically studied the azimuthal asymmetry of
partons in parton-nucleus collisions. The azimuthal asymmetry of
partons in partons-nucleus collisions is very sensitive to the
effective gluon mass and the higher-order radiation corrections.  We
found that the azimuthal asymmetry of gluons $v_2^g$ at small $p_T$
can be negative in $Gp$ and $GA$ collisions, and at higher $p_T$ changes
sign and becomes positive. This is in contrast with the $v_2^q$ of
quarks in $qA$ collisions, where it is positive.

The technique presented in this paper can be also used to study the
azimuthal asymmetry in DIS and in the production of dileptons.
We plan to report on some of these problems in the near future.

\section*{Acknowledgments}
 This work was supported in part by Conicyt (Chile) Programa
 Bicentenario PSD-91-2006, by Fondecyt (Chile) grants 1070517, 1050589
 and by DFG (Germany) grant PI182/3-1.

\appendix
\section{}
\begin{figure}
\includegraphics[height=.20\textheight]{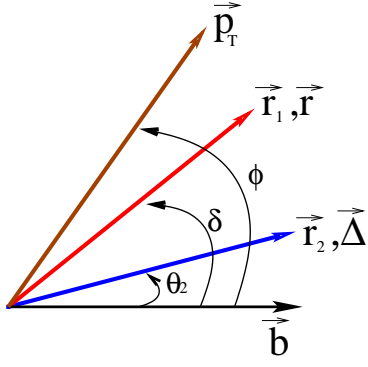}\\
\caption{ The angles defined in Eqs.~(\ref{in2},\ref{f3}). \label{fig7}}
\end{figure}

In this section we illustrate a derivation of Eq.~(\ref{vgn}). At
$\alpha=0$, up to a constant normalization $N$, the averaged product
of the light-cone wave function Eq.~(\ref{100}) reads, \beq
 \overline{\Psi_{GG}^*(\vec r_1,\alpha )\Psi_{GG}(\vec r_2,\alpha )}=
 N \exp\left[-\frac{r_1^2+r_2^2}{2\,r_0^2}\right]\frac{\vec r_{1} \cdot \vec r_2}{r_1^2r_2^2}.
 \eeq
The azimuthal asymmetry is defined as,
\begin{equation}
v_{2}^{g}(p_{T}, b)= \frac{\int_{-\pi}^{\pi} d\phi \cos(2\phi)
\frac{d \sigma (GA\to GGX)}{d^{2}\vec{p}_{T}d^{2}\vec{b}}}
{\int_{-\pi}^{\pi} d\phi \frac{d \sigma(GA\to GGX)}{d^{2}\vec{p}_{T}d^{2}\vec{b}}}
= \frac{I_{N1}+I_{N2}}{I_{D1}+I_{D2}+I_{D3}}, \label{v2-2}
\end{equation}
where the gluon radiation cross section is defined in
Eq.~(\ref{80}). The Fourier-integral is inconvenient for numerical
calculation, but we can perform some of integral analytically in order
to make the numerical task manageable. Let`s define
\begin{eqnarray}
\mathcal{I}_G(b, r,\delta)&=&
\frac{9}{4}\int d^{2}\vec{s}~\text{Im}f^{N}_{q\bar{q}}(\vec{s},\vec{r})T_{A}(\vec{b}+\vec{s})
,\label{I}\
\end{eqnarray}
where the factor $\frac{9}{4}$ comes from the definition Eq.~(\ref{3g-1})
at $\alpha=0$, and $\delta$ denotes the angle between the impact parameter
$\vec b$ and the dipole vector $\vec r$, see Fig.~\ref{fig7}.

We first perform the calculation for the first term in the numerator
and the denominator $I_{N1}, I_{D1}$ in Eq.~(\ref{v2-2}),

\beqn
\begin{pmatrix}I_{N1} \\ I_{D1} \end{pmatrix} &=&-
2\int d\phi d^2r_1d^2r_2  \begin{pmatrix}\cos(2\phi) \\ 1\end{pmatrix}
e^{i\vec p_T(\vec r_1-\vec r_2)} e^{\frac{-r_1^2-r_2^2}{2r_0^2}} \frac{\vec r_{1}
\cdot \vec r_2 }{r_1^2 r_2^2} e^{-\mathcal{I}_G(b, r_1,\delta)}\nonumber\\
&=&-2\int d\phi dr_1dr_2d\delta d\theta_2 \begin{pmatrix}\cos(2\phi) \\ 1\end{pmatrix} \cos(\delta-\theta_2)
e^{i p_T r_1 \cos(\phi-\delta)-\frac{r_1^2}{2r_0^2}-\mathcal{I}_G(b, r_1,\delta)}
e^{-ip_T r_2 \cos(\phi-\theta_2)-\frac{r_2^2}{2r_0^2}}. \label{in2}\
 \eeqn
The angles in the second Eq.~(\ref{in2}) are defined in Fig.~\ref{fig7}.
We can immediately perform the integral over $\theta_2$ in the above equation by using the following identity
\begin{equation}
e^{iA cos(\phi)}=\sum_{n=-\infty}^{+\infty} i^n J_{n}(A) e^{in \phi}, \label{id}
\end{equation}
where $J_{n}(x)$ denotes the Bessel function. We have

\beqn
\mathcal{K}_1&=&\int d\theta_2 \cos(\delta-\theta_2)  e^{-ip_T r_2 cos(\phi-\theta_2)}=\sum_{i=-\infty}^{+\infty} i^n J_{n}(-p_T r_2) e^{in \phi} \int d\theta_2  e^{-in \theta_2}
\cos(\delta-\theta_2)
\nonumber\\
&=&-2\pi iJ_1(p_T r_2)\cos(\phi-\delta).\ \label{k11}
\eeqn
Having plugged the above equation into Eq.~(\ref{in2}), we can then perform the integral over $\phi$,

\beqn
\mathcal{K}_2&=&\int d\phi \cos(\phi-\delta)  \begin{pmatrix}\cos(2\phi) \\ 1\end{pmatrix}  e^{ip_T r_1 cos(\phi-\delta)}=\sum_{i=-\infty}^{+\infty} i^n J_{n}(p_T r_1) e^{-in \delta}\int d\phi
e^{in \phi}  \cos(\phi-\delta)  \begin{pmatrix}\cos(2\phi) \\ 1\end{pmatrix} \nonumber\\
&=&\pi i \begin{pmatrix} \left(J_1(p_T r_1)-J_3(p_T r_1)\right)\cos(2\delta)  \\ 2J_1(p_T r_1)\end{pmatrix}.\ \label{k12}
\eeqn

Now we can simplify Eq.~(\ref{in2}) by using  $\mathcal{K}_1$ and $\mathcal{K}_2$ given in Eqs.~(\ref{k11},\ref{k12}),
\beqn
\begin{pmatrix}I_{N1} \\ I_{D1} \end{pmatrix}
&=&-4\pi^2\int d\delta dr_1 \begin{pmatrix} \left(J_1(p_T r_1)-J_3(p_T r_1)\right)\cos(2\delta)  \\ 2J_1(p_T r_1) \end{pmatrix}   e^{\frac{-r_1^2}{2r_0^2}}e^{-\mathcal{I}_G(b, r_1,\delta)} \int dr_2 J_1(p_T r_2)
e^{\frac{-r_2^2}{2r_0^2}}\nonumber\\
&=&-\frac{(2\pi)^2}{p_T}\left(1-e^{-p_T^2r_0^2/2}\right) \int d\delta dr_1
\begin{pmatrix}\left(J_1(p_T r_1)-J_3(p_T r_1)\right)\cos(2\delta)  \\ 2J_1(p_T r_1) \end{pmatrix} e^{\frac{-r_1^2}{2r_0^2}}e^{-\mathcal{I}_G(b, r_1,\delta)}.\
\label{in1-1}
\eeqn
The remaining integrals in the above expression can be only done numerically. Now we calculate the second term in the denominator and numerator of $v_2^g$ Eq.~(\ref{v2-2}),

\beq
\begin{pmatrix}I_{N2} \\ I_{D2} \end{pmatrix} =
\int d^2\vec r_1d^2\vec r_2 \begin{pmatrix}\cos(2\phi) \\ 1\end{pmatrix}
e^{i\vec p_T(\vec r_1-\vec r_2)} e^{\frac{-r_1^2-r_2^2}{2r_0^2}}\frac{\vec r_1\cdot\vec r_2} {r_1^2 r_2^2}
\exp\left[-\frac{9}{4}\int d^{2}\vec{s}~\text{Im}f^{N}_{q\bar{q}}(\vec{s},\vec{r}_1-\vec r_2)T_{A}(\vec{b}+\vec{s})\right].
 \label{in1}
 \eeq
We first change the variables in the above integrals by:
\beq
\vec r_1=\frac{\vec \Delta+\vec r }{2},   \hspace {2cm} \vec r_2=\frac{\vec \Delta -\vec r}{2}.
 \eeq
Therefore, we obtain,
\beqn
\begin{pmatrix}I_{N2} \\ I_{D2} \end{pmatrix} &=&
\frac{1}{4}\int d\phi d^2\vec r d^2\vec \Delta \begin{pmatrix}\cos(2\phi) \\ 1\end{pmatrix}
 e^{i\vec p_T. \vec r}e^{\frac{-r ^2-\Delta^2}{4r_0^2}}  e^{-\mathcal{I}_G(b, r,\delta)}
\frac{4(\Delta^2-r^2)}{(\Delta^2+r^2)^2-4(\vec \Delta. \vec r)^2},\nonumber\\
&=& \int dr d\Delta d \delta d\phi d \theta_2  \begin{pmatrix}\cos(2\phi) \\ 1\end{pmatrix} e^{i p_T r \cos(\phi-\delta)-\frac{r^2+\Delta^2}{4r_0^2}-\mathcal{I}_G(b, r,\delta)}
\frac{(\Delta^2-r^2)\Delta r}{(\Delta^2+r^2)^2-4(\Delta r\cos(\delta-\theta_2))^2}. \label{f3}\
\eeqn
The angles in the above equation are defined in Fig.~\ref{fig7}. Notice that the factor $\frac{1}{4}$ in the above expression is the Jacobi determinant.
The integral over $\phi$ can be immediately performed using the identity
Eq.~(\ref{id}),

\beq
\mathcal{K}_3=\int d\phi  \begin{pmatrix}\cos(2\phi) \\ 1\end{pmatrix} e^{ip_T r cos(\phi-\delta)}=2\pi \begin{pmatrix} -J_2(p_T r)\cos(2\delta)\\ J_0(p_T r)\end{pmatrix}.
\label{k13}
\eeq
$I_{N2}, I_{D2}$ defined in Eq.~(\ref{f3}) can be then written in the following form
\beqn
\begin{pmatrix}I_{N2} \\ I_{D2} \end{pmatrix} &=&2\pi \int dr d \delta  \begin{pmatrix} -J_2(p_T r)\cos(2\delta)\\ J_0(p_T r)\end{pmatrix}  e^{\frac{-r^2}{4r_0^2}}
e^{-\mathcal{I}_G(b, r,\delta)}f(r,\delta).\
\eeqn
where the function $f(r,\delta)$ is defined,
\beq
f(r,\delta)= \int d\Delta d \theta_2 \frac{(\Delta^2-r^2 )\Delta r}{(\Delta^2+r^2)^2-4(\Delta r\cos(\delta-\theta_2))^2} e^{-\frac{\Delta^2}{4r_0^2}}. \label{ff}
\eeq
The remaining multi-integrals in the above equation can be only carried out numerically.
The last term in the denominator of $v_2^g$ is independent of impact parameter $b$,

\beqn
I_{D3} &=& \int d\phi d^2r_1d^2r_2
e^{i\vec p_T(\vec r_1-\vec r_2)} e^{\frac{-r_1^2-r_2^2}{2r_0^2}} \frac{\vec r_1. \vec r_2}{r_1^2 r_2^2},  \nonumber\\
&=&2\pi \int dr_1dr_2d\theta_1 d\theta_2  \cos(\theta_1)\cos(\theta_2)
e^{i p_T r_1 \cos(\theta_1)-\frac{r_1^2}{2r_0^2}}
e^{-ip_T r_2 \cos(\theta_2)-\frac{r_2^2}{2r_0^2}}. \label{in3}\
 \eeqn
Using the following integral,
\beq
\int d \theta e^{i p_T r_1 \cos(\theta_1)}\cos(\theta_1)=2\pi iJ_1(p_T r).
\eeq
we obtain
\beq
I_{D3} =(2\pi)^3\left(\int dr J_1(p_T r)e^{\frac{-r^2}{2r_0^2}}\right)^2=\frac{(2\pi)^3}{p_T^2}\left(1-e^{-p_T^2 r_0^2/2}\right)^2.
\eeq
To put all equations together, we can  write $v_2^g$ defined in Eq.~(\ref{v2-2}) as $v_2^g=v_2^N/v_2^D$ in a factorized form,
\beqn
v_2^D&=& \frac{(2\pi)^3}{p_T^2}\left(1-e^{-p_T^2r_0^2/2}\right)^2-\frac{2(2\pi)^2}{p_T}\left(1-e^{-p_T^2r_0^2/2}\right)\int dr J_1(p_T r) e^{\frac{-r^2}{2r_0^2}}f_0(b,r)
+(2\pi)\int dr J_0(p_T r)e^{\frac{-r^2}{4r_0^2}}f_1(r,b),\nonumber\\
v_2^N&=&
-\frac{(2\pi)^2}{p_T}\left(1-e^{-p_T^2r_0^2/2}\right) \int dr    \left(J_1(p_T r)-J_3(p_T r)\right) e^{\frac{-r^2}{2r_0^2}}f_2(b,r)
-(2\pi)\int dr J_2(p_T r)e^{\frac{-r^2}{4r_0^2}}f_3(b,r),\nonumber\\
\eeqn
where function $f(r,\delta)$ is defined in Eq.~(\ref{ff})  and $f_{0-3}(b,r)$ in the above equation are defined as follows
\beqn
f_{0}(b,r) &=&\int d\delta  e^{-\mathcal{I}_G(b, r,\delta)}, \nonumber\\
f_{1}(b,r) &=&\int d\delta e^{-\mathcal{I}_G(b, r,\delta)} f(r,\delta), \nonumber\\
f_{2}(b,r) &=&\int d\delta \cos(2\delta)e^{-\mathcal{I}_G(b, r,\delta)}, \nonumber\\
f_{3}(b,r) &=&\int d\delta \cos(2\delta) e^{-\mathcal{I}_G(b, r,\delta)} f(r,\delta).\
\eeqn
In a very similar fashion one can also obtain Eq.~(\ref{vn1}).


\begin{thebibliography}{99}

\bibitem{rhic1}
 K. H. Ackermann {\it et al.}, (STAR Collaboration), Phys. Rev. Lett. {\bf 86}, 402 (2001).
\bibitem{rhic2}
STAR Collaboration, Nucl. Phys. {\bf A757}, 102 (2005).
\bibitem{hadro}
For a review see:  P. F. Kolb and U. Heinz, nucl-th/0305084.
\bibitem{vrhic}
For example:
D. Moln\'ar and S. A. Voloshin, Phys. Rev. Lett. {\bf 91}, 092301 (2003); V. Greco,
C.-M. Ko and P. Levai, Phys. Rev. {\bf C68}, 034904 (2003); R. J. Fries, B. Muller,
C. Nonaka and S. A. Bass, Phys. Rev. {\bf C68}, 044902 (2003).
\bibitem{vrhic2}
D. Molnar and M. Gyulassy, Nucl. Phys. {\bf A697}, 495 (2002), erratum-ibid {\bf A703},
893 (2002).
\bibitem{vvv}
 H.-J Drescher, A. Dumitru, C. Gombeaud and J.-Y. Ollitrault, Phys. Rev. {\bf C76}, 024905 (2007);
 Z. Xu, C. Greiner, H. Stoecker, Phys. Rev. Lett. {\bf 101}, 082302 (2008);
 O. Fochler, Z. Xu, C. Greiner, arXiv:0806.1169. 
\bibitem{me1}
B. Z. Kopeliovich, H. J. Pirner, A. H. Rezaeian and I. Schmidt, Phys. Rev. {\bf D77}, 034011 (2008).
\bibitem{me2}
B. Z. Kopeliovich, A. H. Rezaeian and I. Schmidt, Nucl. Phys. {\bf A807}, 61 (2008).

\bibitem{rhic-vpp}
 STAR Collaboration, Phys. Rev. {\bf C72}, 014904 (2005).
 \bibitem{zues}
S.~Chekanov {\it et al.},  (ZEUS Collaboration),
  PMC Phys.\  {\bf A1}, 6 (2007)
  [arXiv:0708.1478].
 \bibitem{zkl}
  B.~Z.~Kopeliovich, L.~I.~Lapidus and A.~B.~Zamolodchikov,
  JETP Lett.\  {\bf 33}, 595 (1981)
  [Pisma Zh.\ Eksp.\ Teor.\ Fiz.\  {\bf 33}, 612 (1981)].

\bibitem{gbw}
 K. Golec-Biernat and M. W\"usthoff, Phys. Rev. {\bf D60}, 114023 (1999).
\bibitem{me3}
 B. Z. Kopeliovich, A. H. Rezaeian and I. Schmidt, arXiv:0804.2283.

\bibitem{kst2}
B.~Z.~Kopeliovich, A.~Schafer and A.~V.~Tarasov, Phys. Rev. {\bf D62}, 054022 (2000).
\bibitem{r-pion}
S.~Amendolia {\it et al.}, Nucl. Phys. {\bf B277}, 186 (1986).
\bibitem{pom}
R. M. Barnett {\it et al.}, Rev. Mod. Phys. {\bf 68}, 611 (1996).
\bibitem{boris0}
 M. B. Johnson, B. Z. Kopeliovich and A. V. Tarasov,  Phys. Rev. {\bf C63}, 035203 (2001).

\bibitem{spot}
 B. Z. Kopeliovich, I. K. Potashnikova, B. Povh and I. Schmidt, Phys. Rev. {\bf D76}, 094020 (2007).

\bibitem{kst1}
 B. Z. Kopeliovich, A. Schafer and A. V. Tarasov, Phys. Rev. {\bf C59}, 1609 (1999).

\bibitem{boris2}
 B. Z. Kopeliovich, J. Nemchik, A. Schafer and A.V. Tarasov,  Phys. Rev. Lett. {\bf 88}, 232303 (2002).
\bibitem{fields-feynman}
 R. P. Feynman, R. D. Field and G. C. Fox, Phys. Rev. {\bf D18}, 3320 (1978).
\bibitem{own}
For a review see: J. F. Owens, Rev. Mod. Phys. {\bf 59}, 465 (1987).
\bibitem{wang}
For example:  X. N. Wang, Phys. Rev. {\bf C61}, 064910 (2000).
\bibitem{emc}
  M.~Arneodo, Phys. Rept. {\bf 240}, 301 (1994).


\bibitem{lat}
A. DiGiacomo and H. Panagopoulos, Phys. Lett. {\bf B285}, 133 (1992).
\bibitem{ins}
T. Sch\"afer, E. V. Shuryak, Rev. Mod. Phys. {\bf 70}, 323 (1998).

\bibitem{mrst2006}
A. D. Martin, W. J. Stirling, R. S. Thorne, and G. Watt, Phys. Lett. 
{\bf B652}, 292 (2007).
\bibitem{kretzer2000}
S. Kretzer, Phys. Rev. {\bf D62}, 054001 (2000).

\bibitem{ktc2}
 G. Papp, P. Levai and G. Fai, Phys. Rev. {\bf C61}, 021902(R) (1999).
\bibitem{apanasevich98} 
L. Apanasevich {\it et al.}, Phys. Rev. {\bf D59}, 074007 (1999);
\bibitem{ktc3} 
B. Z. Kopeliovich, A. H. Rezaeian, H. J. Pirner and Ivan Schmidt, Phys. Lett. {\bf B653}, 210 (2007);
 A. H. Rezaeian {\it et al.}, arXiv:0707.2040; N. Armesto, (ed.) {\it et al.}, J. Phys. {\bf G35}, 054001 (2008).


\bibitem{rhic2006}
S. S. Adler {\it et al.}, (PHENIX Collaboration), Phys. Rev. Lett.
{\bf 98}, 172302 (2007).

\bibitem{sps79} D. Antreasyan {\it et al.}, Phys. Rev. {\bf D19},
764 (1979).

\bibitem{sps89} D. F. Jaffe {\it et al.}, Phys. Rev. {\bf D40},
2777 (1989).




\end{thebibliography}
\end{document}